\RequirePackage{fix-cm}
\documentclass[smallextended]{svjour3}       
\smartqed  
\usepackage{graphicx,color}
\usepackage{amsmath}
\usepackage{subfigure}   
\begin{document}

\title{Developments in THz range ellipsometry
}


\author{M. Neshat      \and
        N.P. Armitage
}


\institute{M. Neshat \at
              School of Electrical and Computer Engineering, University of Tehran, Tehran, Iran \\
              \email{mneshat@ece.ut.ac.ir}           
           \and
           N.P. Armitage \at
Department of Physics and Astronomy,  Johns
Hopkins University, Baltimore, MD 21218, USA\\
              \email{npa@pha.jhu.edu}     }

\date{\today}


\maketitle

\begin{abstract}
 Ellipsometry is a technique whereby the measurement of the two orthogonal polarization components of light reflected at glancing incidence allows a characterization of the optical properties of a material at a particular frequency.  Importantly, it obviates the need for measurement against a standard reference sample, and so can provide reliable spectroscopic information even when surface morphology is unknown, of marginal quality and/or a reference is unavailable.  Although a standard technique in the visible range, it has not been widely applied in the Terahertz (THz) spectral range despite its potential utility.   This is largely because of the technical difficulties that these frequencies present.   This review details recent progress in the implementation of THz range ellipsometry.  We discuss a variety of configurations including various kinds of laboratory and facility based sources using both continuous wave and pulsed spectroscopic methods.  We discuss the general problems encountered when trying to import the methodologies of visible range ellipsometry to the THz range and give examples of where the technique has been successful thus far.

\end{abstract}


\section{Introduction}
\label{intro}

Ellipsometry is an optical characterization technique in which the change in the polarization state of light upon glancing incidence reflection (or transmission) through a sample is measured \cite{Azzam_1977}-\nocite{Tompkins_1999}\nocite{Tompkins_2005}\nocite{Schubert_2004}\cite{Fujiwara_2007}. Typically, ellipsometry measures the amplitude ratio $\Psi$ and phase difference $\Delta$ between $p-$ and $s-$polarized light. The name \emph{ellipsometry} stems from the fact that the polarization state generally becomes \emph{elliptical} upon light reflection. In spectroscopic ellipsometry, the spectra of the ellipsometric parameters ($\Psi,\Delta$) are measured over a range of wavelengths and can be converted to complex optical constants such as the complex dielectric constant or index of refraction through appropriate models.

Conventionally, spectroscopic ellipsometry measurement has been carried out in the ultraviolet, visible and near infrared regions.  In its optical frequency incarnations it has been used in a wide range of applications \cite{Fujiwara_2007} e.g. in semiconductor industry for analysis of substrates, thin films, and lithography, in chemistry to study polymer films, self-assembled monolayers, proteins, DNA, in the display industry to characterize TFT films, transparent conductive oxides, organic LED, in optical coatings, in real-time monitoring for chemical vapor deposition (CVD), molecular beam epitaxy (MBE), etching, oxidation, thermal annealing, liquid phase processing etc. and for non-destructive testing.

The implementation of spectroscopic ellipsometry in the THz range  (historically called far-infrared or millimeter wave range) has been more problematic.   This part of the electromagnetic spectrum is very important as picosecond ($10^{-12}$ sec) timescales are one of the most ubiquitous in material systems.  Such ubiquity makes spectroscopic tools employing THz electromagnetic radiation potentially very useful. Unfortunately, measurements in the THz spectral range are traditionally challenging to implement as they lie in the so-called ``Terahertz gap'' - above the capabilities of traditional electronics, but below that of optical generators and detectors (photonics).  This is for a number of reasons including weak sources, long wavelengths, and contamination by ambient room temperature black body radiation.  In recent years however, there have been an increasing number of techniques using both lab and facility based (e.g. free-electron laser, synchrotron) that allow measurements that span this gap.  As such, THz spectroscopy has become a tremendous growth field \cite{DOEreport}, finding use in a multitude of areas including characterization for novel solid-state materials \cite{Kaindl,Heyman,Aguilar}, optimization of the electromagnetic response of new coatings \cite{coatings}, probes of superconductor properties \cite{Corson,Bilbro}, security applications for explosives and biohazard detection \cite{THzBiohazard}, detection of protein conformational changes \cite{THzproteins}, and non-invasive structural and medical imaging \cite{THzimaging0,THzimaging1,THzimaging2,Mittleman97a} to give an incredibly incomplete list.

However, an outstanding technical problem in THz spectroscopy remains the difficulty in determining complex optical constants of
materials in reflection mode; generally the only reliable spectroscopic measurements have been on materials that are transparent
enough to perform transmission.   The problems are multi-fold.   Generally in order to generate complex optical constants one must measure or know both a wave's amplitude and phase.  The accuracy of the measurement of amplitude is generally determined by the experimenter's ability to establish the baseline for 100\%  (or some other known value) reflectance.  This is particularly a problem for measuring at low frequencies of conductive materials as the reflectivity of conductors tends to unity as $\omega \rightarrow 0$.

The phase shift of a wave upon reflection is also generally difficult to measure.  For instance in the time-domain THz technique, one measures the time-dependent transmitted electric field of a sample as compared to a reference.  In transmission measurements, transmission through an aperture is used as a reference. For reflection based time-domain THz a simple mirror can not be used as a reference because its surface would need to be positioned within a fraction of a micron of $exactly$ the same place as the sample so that the relative reflected phase is accurate. The general case of sub-micrometer positioning is indeed difficult to achieve, although various methods have been published to avoid this problem (See, e.g. \cite{Nashima_Dec2001}-\nocite{Pashkin_2003}\nocite{Thrane1995330}\nocite{howells:550}\cite{nne:5319}). In techniques that just measure power, one determines the reflected phase by a Kramers-Kronig integral transform rather than by direct measurements.  The results of Kramers-Kronig analysis of reflectivity in the THz part of the spectrum depend heavily on accurate knowledge of the reflectance over a wide energy range and extrapolation beyond it.  The lack of reliable reflection based capabilities for THz measurements has been an impediment to progress in this field.

To solve both the amplitude and phase problem, the technique of spectroscopic ellipsometry has traditionally played an important role.  Ellipsometry obviates the need for measurement against a standard reference sample as it is self-referencing.  Hence it can provide reliable spectroscopic information even when surface morphology is unknown, of marginal quality and/or a reference is unavailable.  Moreover, signal to noise ratios are generally very good, as source fluctuations are divided out.  It is also phase sensitive, as of importance is the phase difference between reflected $s-$ and $p-$ polarized light, which determines the polarization state $a$ $la$ linear, circular, elliptical, etc.

This paper reviews developments in the implementation and usage of \emph{THz ellipsometry}.  Ellipsometry in the THz range has a long history.   Due to the tremendous potential applicability for performing THz measurements in reflection there have been many attempts to establish successful methods of performing such experiments.  A variety of methods and means have been used including various kinds of laboratory and facility based sources using both continuous wave and pulsed spectroscopic methods.  We discuss the general problems encountered when trying to import the methodologies of visible range ellipsometry to the THz range.   We also discuss the various implementations including continuous wave, synchrotron based, and time-domain pulsed methods. In this review we cover only what we consider to be true ellipsometry, e.g. where the $s-$ and $p-$components of either transmitted or reflected $glancing$ incidence light is compared.   For instance, we don't cover recent very good work that has been done with THz light at normal incidence in the form of THz polarimetry on magnetic or magneto-electric materials \cite{Mechelen11a}.

\section{Ellipsometry:   General considerations}
\label{sec:1}

\begin{figure*}[htbp]
\centering
\includegraphics[width=0.6\textwidth]{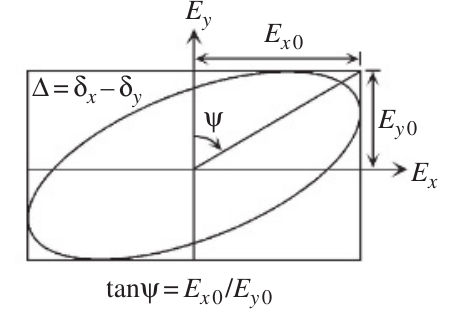}
\caption{Representation of the elliptical polarization by the ($\Psi,\Delta$) coordinate system.}
\label{PSI_DELTA}       
\end{figure*}

The polarization state of elliptically polarized light can be expressed by ($\Psi,\Delta$) parameters as shown in Fig. \ref{PSI_DELTA}.  Here, $\Psi$ and $\Delta$ represent the angles of the amplitude ratio ($\tan\Psi=E_{x0}/E_{y0}$) and phase difference ($\Delta=\delta_x-\delta_y$), respectively.   For polarized light the Jones matrix formalism can be used.   The Jones vector can be written in terms of the ellipsometric parameters as

\begin{equation}
\begin{bmatrix}
E_x\\
E_y
\end{bmatrix}=
\begin{bmatrix}
E_{x0}\exp(i\Delta)\\
E_{y0}
\end{bmatrix}=
\begin{bmatrix}
\sin\Psi\exp(i\Delta)\\
\cos\Psi
\end{bmatrix}.\label{Jones}
\end{equation}

For unpolarized light the more general Mueller matrix formalism must be used.  Here the polarization state of light is expressed by a 4 component Stokes vector.   Incoming and outgoing light are connected by a $4 \times 4$ Mueller matrix $\hat{M}$. For polarized light, the Stokes parameters can also be represented in terms of the ($\Psi,\Delta$) parameters as \cite{Fujiwara_2007}

\begin{subequations}
\begin{align}
S_0&=1,\\
S_1&=-\cos2\Psi,\\
S_2&=\sin2\Psi\cos\Delta,\\
S_3&=-\sin2\Psi\sin\Delta.
\end{align}
\end{subequations}

Additionally Mueller matrix components can be written as a function of the Jones matrix components for optical elements that have no depolarizing effects \cite{Hofmann10a}.  

\begin{figure*}[htbp]
\centering
\includegraphics[width=0.75\textwidth]{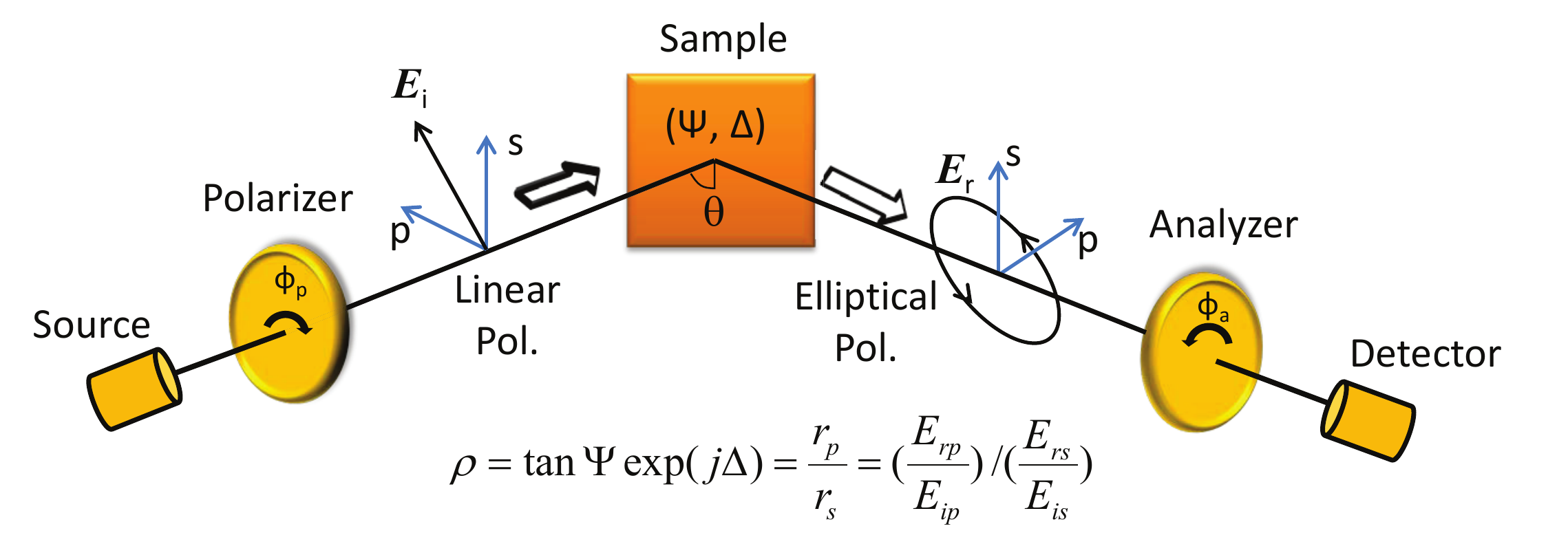}
\caption{Measurement principle of ellipsometry.}
\label{Ellipsometry}       
\end{figure*}

Fig. \ref{Ellipsometry} illustrates the measurement principle of ellipsometry.  In the most common mode of operation - rotating analyzer ellipsometry (RAE) - the source is maintained at fixed polarization, an analyzer polarizer rotates, and the intensity is recorded as a function of angle.  This gives an ellipse that can be characterized in terms of  ($\Psi,\Delta$).  In Fig. \ref{Ellipsometry}, the incident light is linearly polarized at +45$^\circ$ relative to the $p$-axis that results in $E_{ip}=E_{is}$.

The ($\Psi,\Delta$) measured from ellipsometry are defined from the ratio of the
amplitude reflection coefficients for $p-$ and $s-$polarizations:

\begin{equation}
\rho\equiv\tan\Psi\exp(i\Delta)\equiv\frac{r_p}{r_s}\equiv(\frac{E_{rp}}{E_{ip}})/(\frac{E_{rs}}{E_{is}})\label{ratio}
\end{equation}

\noindent where $r_p$ and $r_s$ are defined by the ratios of reflected electric fields to incident electric fields. Eq. (\ref{ratio}) can be simplified to $\tan\Psi\exp(i\Delta)=E_{rp}/E_{rs}$ when $E_{ip}=E_{is}$. Under such condition, $\Psi$ represents the angle determined from the amplitude ratio between reflected $p-$ and $s-$polarizations, whereas $\Delta$ expresses the phase difference between $p-$ and $s-$polarizations.

In general, since the direct interpretation of the ellipsometric parameters is difficult, the construction of an optical model is necessary for data analysis. Once $\Psi$ and $\Delta$ spectra are obtained from experiments, the optical constants and layer thicknesses can be generally extracted by fitting proper dielectric functions and optical models. For anisotropic uniaxial materials the measured $\rho$ is directly related to the so-called \textit{pseudo-dielectric} function as

\begin{equation}
\epsilon =\mathrm{ sin^2}\theta\Big[1 + \mathrm{ tan^2} \theta \big(\frac{1 - \rho}{1+\rho}\big)^2\Big].
\end{equation}

The role of anisotropy and the relation between the pseudo-dielectric and true tensorial dielectric function are discussed below.   In the specific case of a bulk homogeneous sample the actual complex refractive index ($n+ik$) can be directly obtained from the ellipsometric parameters by using the expressions \cite{Born_1999}

\begin{align}\label{bulk_Eq}
n^2-k^2&=\sin^2\theta\bigg[1+\frac{\tan^2\theta(\cos^22\Phi-\sin^22\Phi\sin^2\Delta)}{(1+\sin 2\Phi\cos\Delta)^2}\bigg],\\\nonumber
2nk&=\sin^2\theta\frac{\tan^2\theta\sin 4\Phi\sin\Delta}{(1+\sin 2\Phi\cos\Delta)^2},
\end{align}

\noindent where $\theta$ is the incidence angle, and $\tan\Phi=1/\tan\Psi$.    The dielectric constant follows from the standard relation $\epsilon = (n+ik)^2$.

The method of RAE has broad applicability for isotropic samples with the magnetic permeability $\mu \approx 1$.  In the case of extremely anisotropic sample $s-$ and $p-$polarizations are not the eigenpolarizations of reflection.   In this case or in cases where $\mu \neq 1$ the Jones reflection matrix has  off-diagonal terms,

\begin{equation}
\hat{R} = \left[\begin{array}{cc}r_{ss} & r_{sp} \\r_{ps} & r_{pp}\end{array}\right]. 
\end{equation}

In materials which are magnetoelectric e.g. where electric polarization is caused by magnetic fields and vice versa one can use the the $6\times6$ Berreman matrix equation to express the electromagnetic response \cite{Berreman_Apr1972},
 
 \begin{equation}
\left[\begin{array}{cc}0 & \nabla \times \\   \nabla \times & 0 \end{array}\right] 
\left[\begin{array}{c}E\\H\end{array}\right]  = i \frac{\omega}{c} \left[\begin{array}{cc} \epsilon & \alpha   \\ \alpha' & \mu \end{array}\right]\left[\begin{array}{c}E\\H\end{array}\right]. 
\end{equation}

Here $\epsilon$ and $\mu$ have their usual meanings while $\alpha$ and $\alpha'$ are magneto-electric coefficients in the manner $D = \epsilon E + \alpha H$ and $B = \mu H + \alpha' E$.   For pure transverse waves it is possible to reduce the $6\times6$ Berreman matrix equation to a $4\times4$ matrix equation that suppresses the $z$ component of the electromagnetic fields.

In order to measure anisotropic,  magnetic ($\mu \neq 1$), or magneto-electric ($\alpha \neq 0$) materials within a RAE approach, multiple angles of incidence and sample orientations must be measured.   A more efficient approach in these cases is Muller matrix spectroscopic ellipsometry (MM-SE), where the sample's glancing reflectance is obtained using a number of  linearly-independent states of polarization.   The MM-SE formalism is based on the Stokes vector representation for the light polarization where the real components of the 4D Stokes vectors for the input $S_{in}$ and output $S_{out}$ light are connected with the $4 \times 4$ Mueller matrix $\hat{M}$.  Frequently a normalized Mueller matrix  $\hat{M}/M_{11}$ is used that has 15 independent real components.  As a completely unrestricted $4\times4$ Berreman matrix has 16 complex components, the use of symmetries are required to decouple $\epsilon$, $\mu$ from $\alpha$, $\alpha'$ using the 15 real components available in the normalized Mueller matrix.

In ellipsometry, the incidence angle is typically chosen to be close to the Brewster's angle to maximize the measurement sensitivity as the $s-$ and $p-$ reflection coefficients are maximally different there.  Obviously then the choice of the incidence angle, depends on the optical constants of samples e.g. for semiconductor characterization, the incidence angle is typically 70-80$^\circ$. Multilayer samples with several unknown parameters should also be characterized with multiple incidence angles. Therefore, it is important for ellipsometry systems to have the capability of providing variable incidence angle. It is notable that for isotropic samples at normal incidence the distinction between $p-$ and $s-$polarization is not possible, therefore, ellipsometry at normal incidence becomes impossible for such samples.

Spectroscopic ellipsometry instruments require various error corrections. The sources of error generally come from imprecisions in the zero-degree positions of optical elements that have to be adjusted accurately to the coordinates of $p-$ and $s-$polarizations, leakage in the polarizing elements, non-ideality of the source and detector, and depolarization effects of samples. There is a wealth of knowledge in the literature to compensate for these errors through various calibration schemes \cite{Collins_1990,Jellison_1997,deNijs_1988,Johs_1993,Nguyen_1991,Roseler90a,Ferrieu89a,Bremer92a}.   Their specific use in THz range ellipsometry will be discussed below.

\section{Challenges in implementation of THz ellipsometry}

Although the general scheme for performing THz range ellipsometry can be imported directly from what is known in the UV/VIS range, the THz/FIR range presents a number of special challenges.   Some of these parallel the challenges of implementing THz range spectroscopies in general:  weak sources, large beam sizes, and contamination by ambient black body radiation.   Other of these challenges are more specific to ellipsometry.

One specific such issue is the one of focussing.  In general, analysis of ellipsometry data involves essentially the inversion of the Fresnel equations.   Therefore exact knowledge of the angle of incidence is very important. The effect of uncertainty in the angle of incidence in conventional ellipsometry has been discussed in Ref. \cite{Aspnes85a}.  In principle, a significant source of uncertainty in this regard can be the focussing opening angle.  Even in conventional UV/VIS ellipsometry, there are trade-offs with how tightly one focuses light.  Tight focussing gives a larger opening angle which introduces errors in the optical constants, but gives a small beam spot in which one can use smaller samples.   The need for `large' samples is a fundamental constraint in ellipsometry as measurements are done at glancing incidence.  If one just uses a small aperture in an attempt to maintain a small beam spot, with a small focus opening angle, this actually results in counterproductive unacceptable beam divergences due to diffraction except if the wavelength of light is short.  The problems in this regard in the THz range are potentially even more acute since wavelengths of light are large and light diffracts strongly unless it is strongly focussed or the beam spot is large.   In the first case the angle of incidence may be considered to be poorly defined and in the second only large samples can be used.

\begin{figure}
\centering
  \includegraphics[width=8 cm]{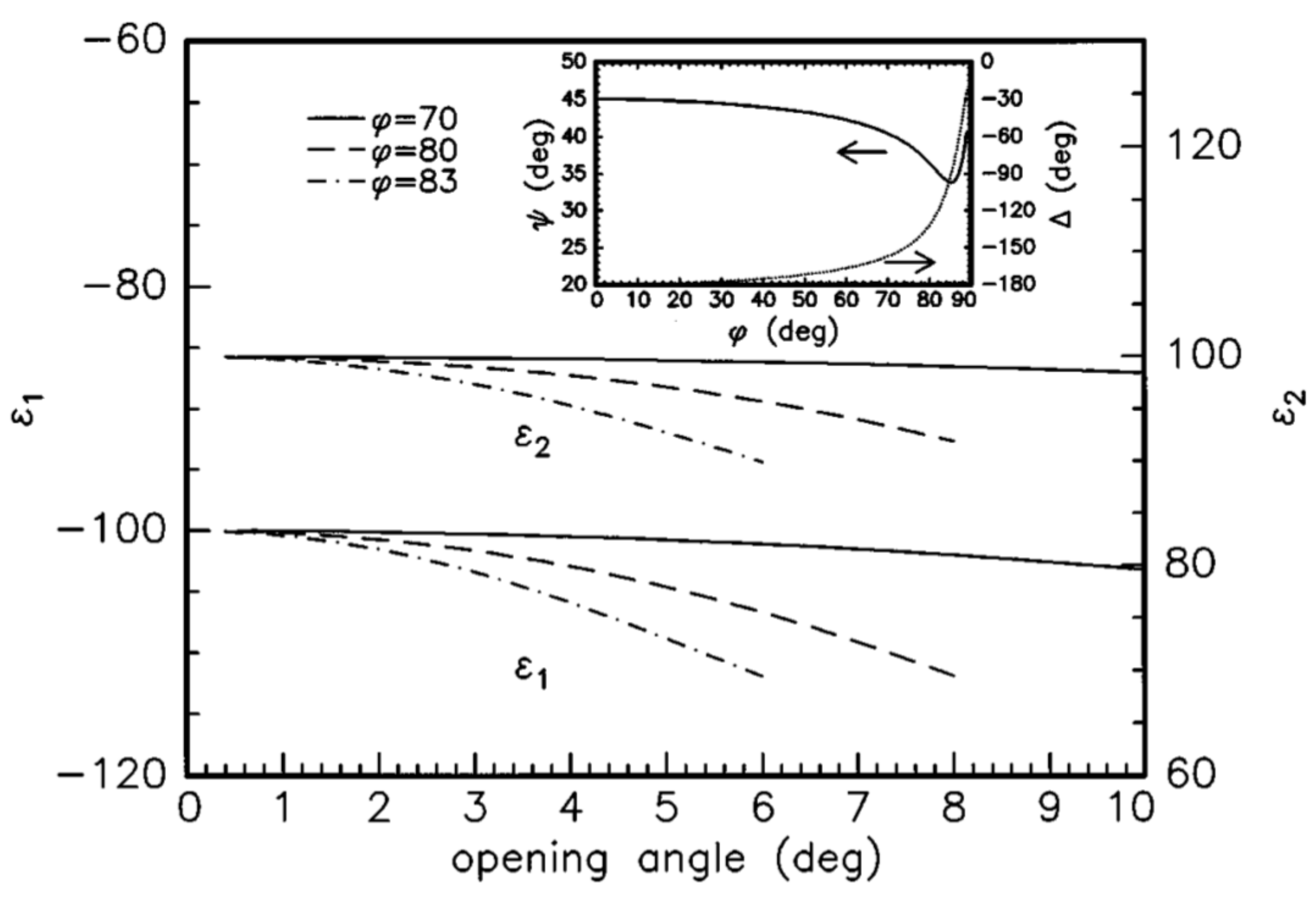}
\caption{Simulation of the measured dielectric function $\epsilon$ as a function of the opening angle $\Delta \varphi$. The angles of incidence are 70$^\circ$ (solid curves), 80$^\circ$	(dashed - curves)	and	83$^\circ$	(dashed - dotted curves).	The inset shows $\Psi$ and $\Delta$ as a function of the angle of incidence $\varphi$.  All data presented in this figure assume a sample with a dielectric constant of $\epsilon = - 100 + i 100$.  From Ref. \cite{Kircher97a}}
\label{KircherSimulation}       
\end{figure}

There have been two main routes to overcome these issues.   The first one historically which was pursued in ellipsometry efforts by Cardona and collaborators was to limit the opening angle and either use only large samples and/or use very intense sources (e.g. synchrotrons).  Kircher \textit{et al.} \cite{Kircher97a} argued that focussing opening angles of more than 5$^\circ$ lead to unacceptable errors if more sophisticated data treatments are not performed.   For a sample with a dielectric constant of $-100 + i 100$ they simulated a number of opening angles at different angles of incidence (Fig. \ref{KircherSimulation}) in a ray optics formalism.   They found that for 70$^\circ$ angle of incidence even a 10$^\circ$ opening angle was tolerable, but for 83$^\circ$ and above, an opening angle of even 5$^\circ$ yields relative errors in the dielectric constant of the order of 10\%.  This can be understood by looking at the inset to Fig. \ref{KircherSimulation} which plots $\Psi$ and $\Delta$ as a function of the angle of incidence $\varphi$.    In general  $\Psi$ and $\Delta$ are slowly varying except near the Brewster angle (which is again why this is the typical angle of incidence for performing ellipsometry).   So within a ray optics formalism one can minimize the errors associated with finite opening angles by working away from the Brewster's angle, but then the precision of the technique falls.  It was suggested by Kircher \textit{et al.} \cite{Kircher97a} that these errors could possibly be dealt with by sophisticated data analysis routines that account for the averaging over opening angles.   Ultimately, however, these authors felt that due to the constraints on beam size in the THz range that the only way to deal with it would be to work with very intense sources.   Their and following attempts to use synchrotrons in this regard is detailed below.

In contrast to this approach, it is argued by the present authors that tight focusing is not as detrimental to the actual effective opening angle as one would expect in a ray optics approach due to diffraction effects in long wavelength THz measurements.  A similar argument has been made in Ref. \cite{Matsumoto11a}.  In the Gaussian approximation, the beam width is given by the expression $w = w_0 \sqrt{1 + (z/z_R)}$ where $z_R$ is the so-called Rayleigh length.  Writing the derivative $dw/dz$ in terms of $\theta = \lambda / \pi w_0$ the opening angle given by ray optics, the wavelength, and using the expression for the Rayleigh length $z_R = \lambda / \pi \theta^2$ one has for the effective angle made by the phase front of the incident wave as related to the angle of incidence

\begin{equation}
\phi = \frac{z  \pi \theta }{\lambda} / \sqrt{1 + z^2 \pi^2 \theta^4 / \lambda^2}.
\end{equation}

If the opening angle $\phi$ must be below the 5$^\circ$ found by  Kircher \textit{et al.} \cite{Kircher97a}, one finds that for a tight geometric opening angle $\theta$ of 26$^\circ$ (used in the terahertz
time-domain spectroscopic ellipsometry (TDSE) setup below) that the sample surface must lie within $\pm$2.5 mm of the geometric focus to maintain this condition for at least the 300 GHz light.   

Another challenge in the THz range is the efficiency of the optical components.  Ellipsometry requires the use of polarized light.   Ellipsometry in this range is hampered by the fact that the best linear polarizers in this spectral range made from free-standing wire grids are still not very good with extinction ratios approaching 1:40.  It is also important not only that the polarizers provide a high degree of polarization, but also that their lateral homogeneity is very good, since the cross-section of the THz beam is typically not very well defined and a spurious intensity modulation can be introduced upon rotating the analyzer polarizer.

\section{Review of previous THz ellipsometry }

\subsection{Far infrared ellipsometry using continuous wave sources}

Conventional ellipsometry in the visible and ultraviolet range typically uses continuous wave sources where the measurement frequency is sequentially stepped.   Early attempts to implement THz range ellipsometry tried either to use this method of sequential stepping a continuous wave source or using a continuous, but broad band source and taking advantage of the frequency multiplexing afforded by Fourier Transform Infrared (FTIR) spectroscopy.  Following work in the mid-infrared by R\"{o}sler \textit{et al.} \cite{Roseler90a} and others \cite{Ferrieu89a,Bremer92a}, Barth \textit{et al.} \cite{Barth93a} constructed an FIR spectroscopic ellipsometer based on a commercial Bruker 113v FTIR instrument.  The spectral range covered was 0.9 - 18 THz.  A number of technical issues had to be confronted in this early attempt that turn out to be important for all attempts at FIR/THz ellipsometry.  Their beam had a convergence angle of about 7$^\circ$, but an aperture system allowed them to reduce this angle to below 0.8$^\circ$. The resulting spot size of the beam at the focus was about 1 cm.   As discussed above, ellipsometric measurements always require knowledge of the plane of incidence as well as the exact zero positions of the polarizer and the analyzer with respect to this plane. They calibrated their polarizer positions essentially $ex$ $situ$ by aligning against the diffraction beam created by a HeNe laser and claim to have gotten precision of 0.1$^\circ$.  The analyzer was rotated stepwise and remains stationary until the data acquisition is finished at each point. Typically, spectra were taken in steps of 9$^\circ$, from 0$^\circ$ to 171$^\circ$, converted to angle-dependent intensity curves, and analyzed in the usual way to yield $\Psi$ and $\Delta$ and then the complex $\epsilon$.  They performed measurements on both Si:P and YBa$_2$Cu$_3$O$_7$ films and obtained reasonable results for the pseudo dielectric function and data that was in good agreement with data generated by Kramer$s-$Kronig analysis of normal-incidence reflectance spectra.

Schubert \textit{et al.} \cite{Schubert03a} demonstrated generalized ellipsometry at far-infrared wavelengths (150 cm$^{-1}$ - 600 cm$^{-1}$) for measurement of the anisotropic magneto-optical response of the semiconductor layered structure i-GaAs/n-GaAs with an external magnetic field.  The ellipsometer operated in the rotating analyzer ellipsometer configuration with an adjustable polarizer.  The focal length of the mirror, which directed the incidence beam onto the sample, was 190 mm, with a typical opening angle of 3$^\circ$.  A Nicolet Fourier transform infrared spectrometer was used for the light source and detector.  They determined a truncated $3\times3$ upper left block of the normalized MM and subsequently determined the frequency dependent magneto-optical dielectric function tensor elements of the n-type GaAs substrate.  It was shown that the band mass, mobility $\mu$, and density could be determined from fits of dielectric tensor element spectra to the Drude model.  The full Muller matrix (MM) was not accessible in this work due to the lack of suitable broad band waveplate to generate cicurlarly polarized light.

Hofmann \textit{et al.} \cite{Hofmann10a} reported an experimental setup for frequency-domain ellipsometric measurements in the 0.2 to 1.5 THz spectral range employing a benchtop-based backward wave oscillator source (BWO).  The instrument allowed angles of incidence between 30$^\circ$ and 90$^\circ$ and operates in a polarizer-sample-rotating analyzer scheme. Their backward wave oscillator source has a tunable base frequency of 107-177 GHz, but with a set of Schottky diode frequency multipliers the spectral range could be extended to 1.5 THz.  The system was calibrated using the regression calibration method for a rotating analyzer element ellipsometer \cite{Johs_1993}.  A model that exactly describes the response of the ideal rotating analyzer polarizer was fit to the data to determine the polarizer and analyzer azimuthal offsets simultaneously with the ellipsometric parameters.  The advantage of this is that the calibration is $in$ $situ$ and does not require a specific known reference sample or an external calibration.  Hofmann \textit{et al.} \cite{Hofmann10a} showed that when applied to P-doped Si at room temperature their ellipsometric data can be well described within the classical Drude model and in excellent agreement with mid-infrared ellipsometry data obtained from the same sample for comparison.  This system was also used to determine the free charge-carrier diffusion profiles in a  $p-p^+$ silicon homojunction \cite{Hofmann09a}.  These authors have also used this system to determine the room-temperature free-charge carrier mobility, sheet density, and resistivity parameters in epitaxial graphene.  They found good agreement between carrier concentration and mobility parameters determined using THz-optical Hall effect (transmission in perpendicular magnetic field) and electrical Hall effect techniques \cite{Hofmann11a,Hofmann11b}.

\begin{figure}
\centering
  \includegraphics[width=7 cm]{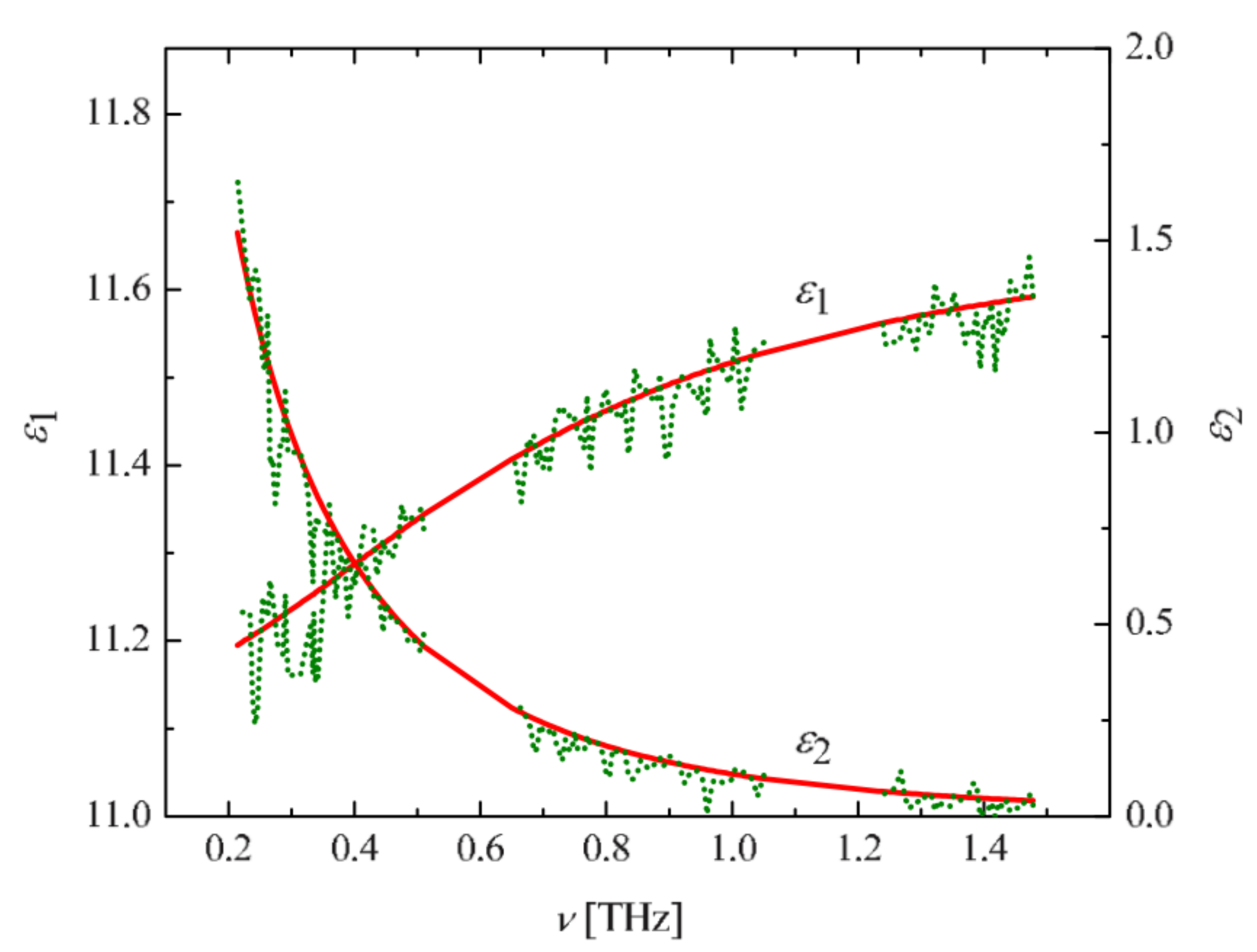}
\caption{Measured (dotted lines) and Drude model fit real ($\epsilon_1$) and imaginary ($\epsilon_2$) part of the frequency dependent dielectric function of the low P-doped $n$-type silicon substrate.  From Ref. \cite{Hofmann10a}.}
\label{SiP}     \end{figure}


\subsection{Far infrared ellipsometry using synchrotron based sources}

To overcome the issues detailed above regarding the limitations imposed from beam size on the size of the sample, successful attempts have been made to use the extremely bright radiation from synchrotron radiation sources.  This overcomes the throughput issues endemic to using small samples as the brilliance of synchrotron radiation in the THz range can be 3 orders of magnitude higher than for thermal black body radiation.  Most of these spectrometers couple radiation from the synchrotron through the optical setup into a commercial FTIR setup for spectroscopic analysis.   The first such setup was at the U4IR beam line of the National Synchrotron Light Source (NSLS), Brookhaven National Laboratory \cite{Kircher97a}.  This setup was equipped with a rotating analyzer and a stationary retarder.  Early experiments were performed on heavily doped GaAs samples ($\approx 8 \times 10^{18}/cm^3$), the ionic insulator LiF, La$_{2-x}$Sr$_x$CuO$_4$, and YBa$_2$Cu$_3$O$_7$.  In these experiments all samples were measured at an angle of incidence of 80$^\circ$ with a beam divergence of $\pm1.2^\circ$ ($f/24$).   There was good agreement with expectations down to frequencies as low as 2 THz.   For instance, they were able to reproduce the well-established optical response of n-doped GaAs and LiF.  Furthermore, they  measured the c-axis response of the high-temperature superconductors La$_{2-x}$Sr$_x$CuO$_4$ and YBa$_2$Cu$_3$O$_7$ at room temperature and determined the phonon parameters of all modes (including the atomic dynamic effective charges in the La$_{2-x}$Sr$_x$CuO$_4$ case) observed in this polarization.

\begin{figure}
\centering
  \includegraphics[width=6 cm]{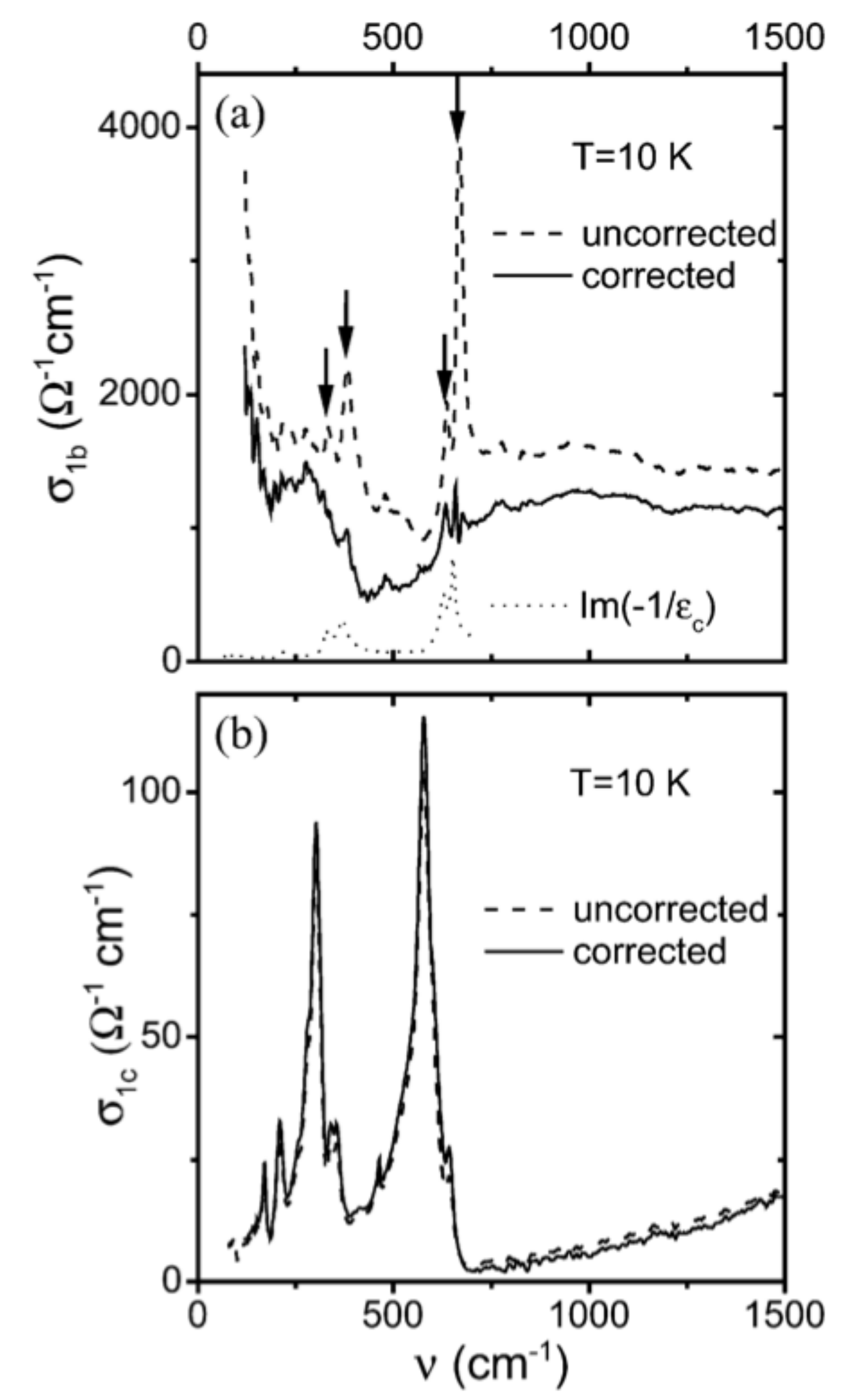}
\caption{Effect of the anisotropy on the measured optical conductivity of  T$_c = $ 91K Bi$_2$Sr$_2$CaCu$_2$O$_{8+ \delta}$.  (a) and (b)  The real part of the conductivity derived from the ellipsometric data before (dashed lines) and after correction for anisotropy effects (thick solid lines), along the the b-axis in-plane component and along the c-axis.  Also shown by the dotted line in (a) is the loss function -Im[$1/\epsilon_c$] of the c-axis component that exhibits peaks near the frequencies of the longitudinal optical phonons.  The features in the in-plane conductivity that arise due to the anisotropy are marked by arrows. From Ref. \cite{Bernhard01a}.}
\label{Bi2212}     \end{figure}

Subsequently a similar setup was built at ANKA at Forschungszentrum Karlsruhe  \cite{Bernhard04a} which works down to $\approx$ 100 cm$^{-1}$.   This setup uses a very small opening angle of  $1^\circ$ and again couples radiation $via$ a series of mirrors through a commercial Bruker FTIR spectrometer on to the sample and then to a Si-based IR bolometer detector.  In this setup it is desirable to introduce a phase shift between $s-$ and $p-$ polarizations because otherwise nearly linearly polarized light incident on the rotating analyzer would mean that almost zero intensity would be measured at certain analyzer angles.   This was done with a pure Si prism that introduces a 84$^\circ$ shift between polarization components.   This setup has been shown to be most useful in the range of 100 cm$^{-1}$ to 750 cm$^{-1}$.  It has been used on a variety of materials including high-temperature cuprate superconductors Bi$_2$Sr$_2$CaCu$_2$O$_{8+ \delta}$  and YBa$_2$Cu$_3$O$_7$   \cite{Bernhard04a} and played a substantial role in the discussions about the possibility of a kinetic-energy driven mechanism of superconductivity in these compounds \cite{Boris04a}.   In the cuprates only modest corrections are necessary to correct the measured pseudo-dielectric function due to c-axis anisotropy.   These corrections are largest on the most anisotropic cuprate Bi$_2$Sr$_2$CaCu$_2$O$_{8+ \delta}$.  In these anisotropy corrected spectra (Fig. \ref{Bi2212}), they have identified a number of features that derive from IR-active in-plane polarized phonon modes.  These have anomalously large spectral weight that they argue to derive from coupling to charge carriers \cite{Bernhard01a}.   They also found a broader feature which they associate with a collective electronic mode around 10 THz, that they speculate corresponds to the pinned phase mode of a charge-density wave.

An FIR range ellipsometer based on a Martin-Puplett interferometer has been built at BESSY \cite{Roseler05a}.  Martin-Puplett interferometers are in general superior in the THz spectral range below 3 THz due to their higher resolution and greater sensitivity over that of standard Michelson-type interferometers.  All Stokes parameters of polarized light can be determined without the need of a retarder as a phase shifting device. This setup has been tested below 1.5 THz using a bending magnet as a dipole radiation source.  BESSY can deliver THz radiation in two different coherent modes where the brilliance of the coherent radiation increases by 3 to 5 orders of magnitude over the incoherent case.

\begin{figure}
\centering
  \includegraphics[width=10cm]{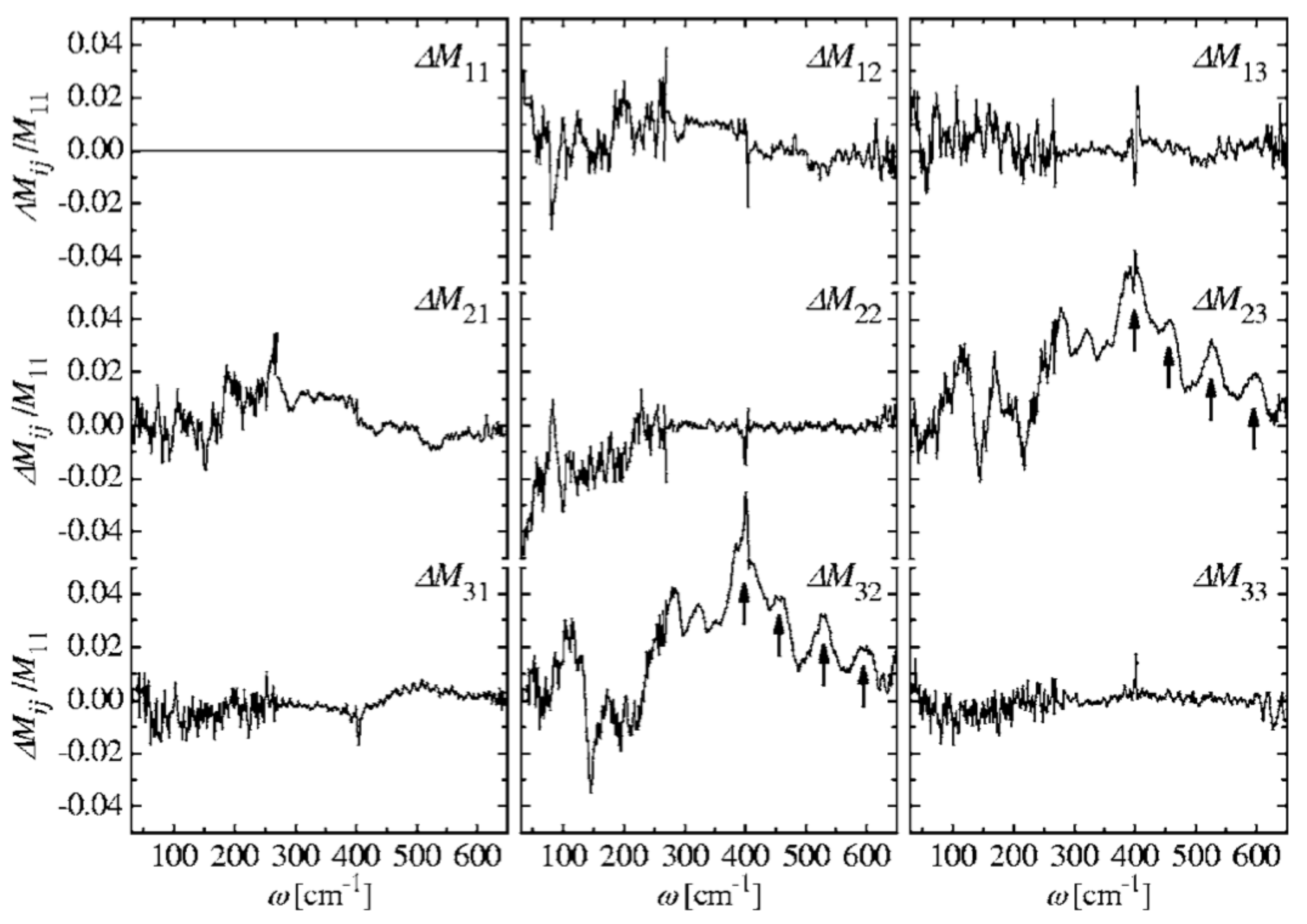}
\caption{Experimental difference spectra (solid lines) of measured Mueller matrix elements M$_{12}$, M$_{13}$, M$_{21}$, M$_{22}$, M$_{23}$, M$_{13}$, M$_{32}$, M$_{33}$ for graphite.  The difference spectra are obtained as the difference between fields $\mu_0 H = 4$ T and 0 T in order to emphasize the field induced changes on the Mueller matrix elements.  The field is actually applied in the incoming beam direction so the field in the $c$ direction is 2.8 T.  From Ref. \cite{Hofmann06a}.}
\label{Graphene}       
\end{figure}

\begin{figure}
\centering
  \includegraphics[width=6cm]{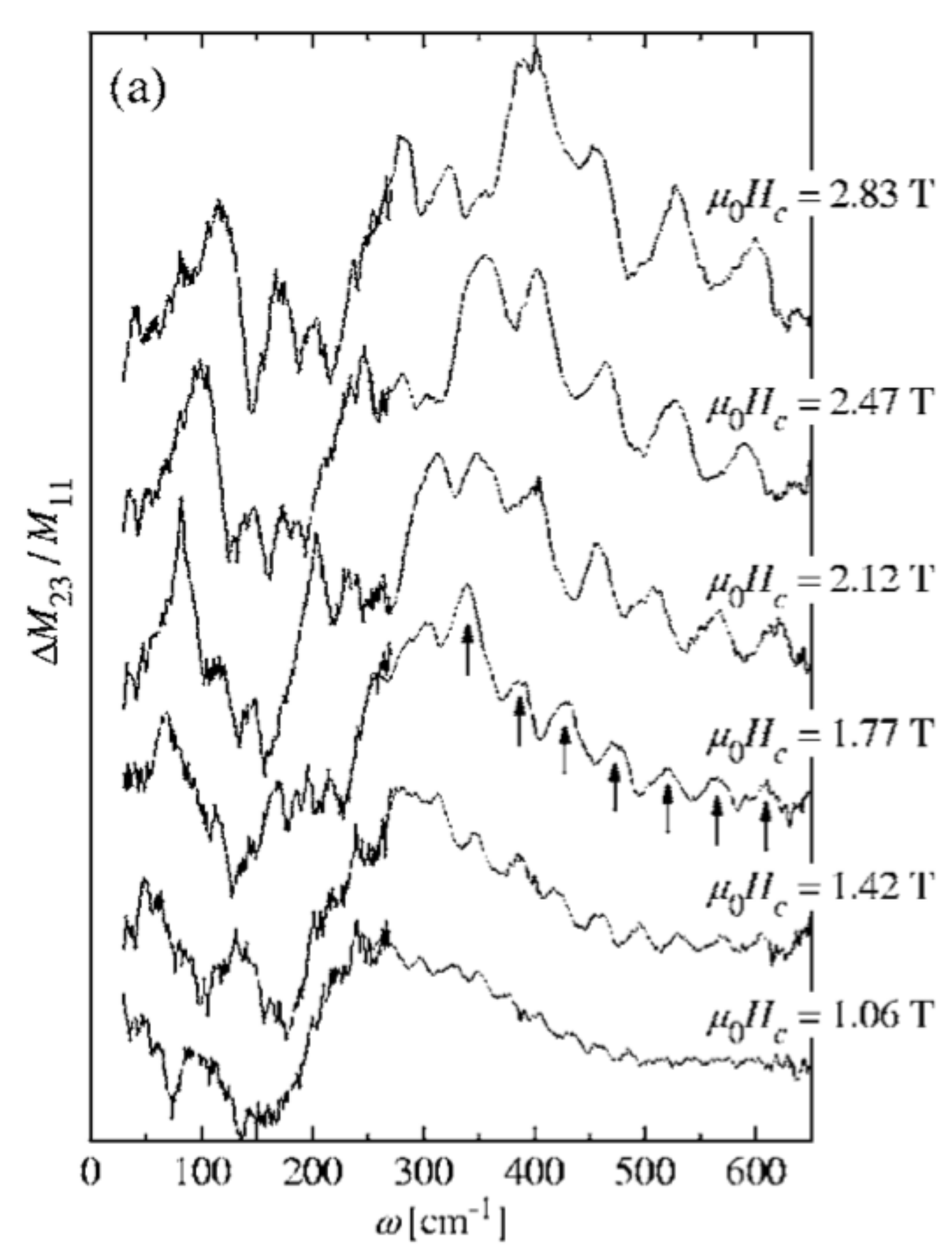}
\caption{The experimental difference spectra of the Muller matrix element M$_{23}$ normalized to M$_{11}$.  The component of the magnetic field H$_c$ parallel to the graphite $c$ axis is given.   Note that these difference spectra are between data at $H_c = \pm x$ T.  From Ref. \cite{Hofmann06a}.}
\label{Graphene2}       
\end{figure}

The first ellipsometer setup capable of performing measurements in the external magnetic field using synchrotron radiation was described by T. Hofmann \textit{et al.} \cite{Hofmann06a,Hofmann06b}. This ellipsometer could measure in both RAE and MM mode where in the latter case, truncated measurements of the $3 \times 3$ upper left block of normalized MM in the spectral region from 30 (0.9 THz) to 650 cm$^{-1}$ (19.5 THz) were possible.  The first example they gave of the kinds of measurements one could perform is of low-chlorine-doped ZnMnSe, a dilute magnetic semiconductor.  Analysis of the normalized Mueller matrix elements using the Drude magneto-optic dielectric function tensor model allowed the determination of the free-charge-carrier properties effective mass, concentration, and mobility over the entire of spectral range of the spectrometer.  They also present Mueller matrix spectra obtained from highly oriented pyrolytic graphite at low temperatures and find signatures of chiral electronic transitions between Landau levels.  The electronic states of quasi-2D metallic systems split into these discrete energy states when subjected to external magnetic fields.  The prominent structures seen in the M$_{23}$ and M$_{32}$ spectra in Fig. \ref{Graphene}, can be attributed to transitions between Landau level transitions separated by multiple amounts of the cyclotron frequency $\omega_c$.  The M$_{11}$ spectrum is flat because of the normalization of all matrix elements to M$_{11}$.  In Fig. \ref{Graphene2}, one can see the Landau level structure which increase in amplitude and period with increasing field, and correspond to circularly polarized Landau level transitions.

Recently a new ellipsometer has been installed at the U4IR beamline of the NSLS \cite{Stanislavchuk13a,Rogers11a}.   This beamline is situated at a dipole bending magnet and is extracted from the storage ring through an aperture that collects an angular range of 90 mrad $\times$ 90 mrad.  NSLS is one of the brightest synchrotron light sources in the world in the far-IR spectral range.  A series of optics collimates the beam and transports it to a Bruker IFS 66v FTIR spectrometer to span the spectral range from far-IR to visible light.   The design of this instrument also takes the approach that in order to get a well-defined angle of incidence one must maximize the $f$-number (greater than 20 in this case).  In this case then the diffraction limited spot is 50$\lambda$ and the instrument's throughput is diffraction limited at 1.5 THz for a 1 cm sized sample.   A large $f$-number was also desired to minimize depolarization effects from the linear polarizers as they are not placed in foci of the optical setup.   A very detailed description of this setup can be found in Ref. \cite{Stanislavchuk13a}.

This spectrometer can work in both conventional rotating analyzer ellipsometry (RAE) and Mueller matrix spectroscopic ellipsometry (MM-SE) modes.  To the best of our knowledge it is the only THz range system to demonstrate MM-SE to determine the full $4\times4$ Mueller matrix.  A challenge in the implementation of MM-SE in the THz regime has been in obtaining broadband circularly polarized light as broadband retarders for this spectral range are not commercially available. The BNL group has developed rotating retarders based on triangular prisms or double Fresnel rhombs  \cite{Kang11a}.  Multiple constraints had to be met in this design, such as a uniform $\pi/2$  phase difference between two orthogonal linear polarization components across the spectral range and a small lateral displacement of the beam when the retarder rotates around its axis.

This beamline has been used to investigate a number of magnetic and magneto-electric compounds.  Dy$_3$Fe$_5$O$_{12}$ garnet is a ferrimagnetic material with a huge magnetostriction that is related to the combination of a strong anisotropy of the crystal field of the RE$^{3+}$ ions and a strong and anisotropic superexchange interaction between the RE and iron.   Its space group is $Ia3d(O_h^{10})$, which does not break inversion symmetry, so the off-diagonal  $\alpha$ blocks in the Berreman matrix must be zero.  Rogers \textit{et al.} \cite{Rogers11a} did observe several hybrid modes with a mixed magnetic and electric dipole activity (e.g. both $\epsilon \neq 1$ and $\mu \neq 1$).   These modes, which arise principally from crystal field excitations, derive their mixed character from their spectral proximity to the phonon at 81 cm$^{-1}$, local electric polarization, and the non-collinear spin structure for the Dy-Fe magnetic system.

\begin{figure}
\centering
  \includegraphics[width=8 cm]{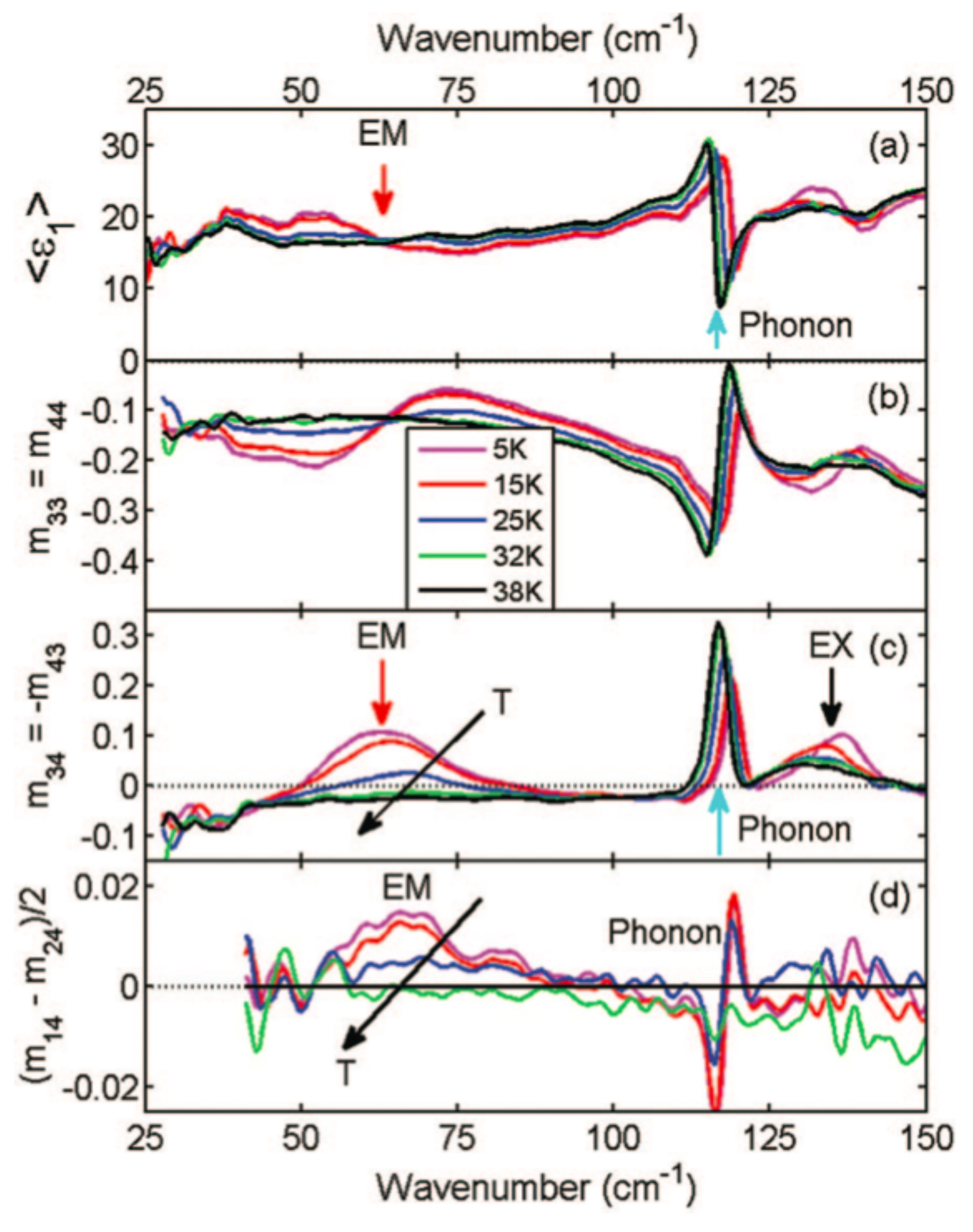}
\caption{Ellipsometric spectra of TbMnO$_3$ with crystallographic orientation of $x||a$, $y||c$, and $z||b$ measured in the temperature range between 5 K and 40 K at the U4IR beamline of the NSLS.  The phonon peak at 118.5 cm$^{-1}$, the electromagnon (EM) peak at 62 cm$^{-1}$, and a peak of unknown origin at 135 $^{-1}$ (EX) are marked with arrows. (a) Real part of the pseudo-dielectric function. (b) The normalized MM components  $m_{34}(\omega) = - m_{43}(\omega)  $ and (c) $m_{44}(\omega) = m_{33}(\omega) $ . (d) The difference between the off-diagonal MM components $ m_{14}(\omega) $ and $m_{24}(\omega)$. Vertical arrows in (c) indicate the contribution of $\epsilon_{xx}(\omega)$, $\mu_{yy}(\omega)$, and $\alpha_{xy}(\omega)$ to $m_{34}(\omega) = - m_{43}(\omega)  $ components of the MM spectra. From Ref. \cite{Stanislavchuk13a}}
\label{TbMnO3}       
\end{figure}

TbMnO$_3$ is a perovskite that has a low-temperature structure that allows both electric polarization and magnetic order.   These orders break both inversion and time-reversal symmetry allowing a non-zero magneto-electric component $\alpha$ \cite{Pimenov06a,Sushkov07a,Aguilar09a}.   It has been proposed that mixed excitations, so-called electro-magnons, can arise in these compounds where the time-varying electric field can excite a transient magnetic response and vice-versa.  It was shown \cite{Pimenov06a} that at frequencies below 100 cm$^{-1}$ (3 THz) the electric component of light directed along the $a$ axis can excite both electromagnons at 19 cm$^{-1}$ (0.57 THz) and 62 cm$^{-1}$ (1.86 THz).  The real part of pseudo-dielectric function is shown in Fig. \ref{TbMnO3}a, which was obtained from RAE spectra measured at temperatures between 5 K and 38 K.  It shows a re-distribution of the spectral weight from the optical phonon at 100 cm$^{-1}$ (3 THz) towards the electromagnon at 62 cm$^{-1}$ , and unknown excitation at 135 cm$^{-1}$ which is labeled as EX.   The spectral weight transfer occurs primarily below the ferroelectric transition temperature of T$_c$ = 28 K.    The normalized MM components are shown in Figs. \ref{TbMnO3}b and \ref{TbMnO3}c.  The difference between the normalized off-diagonal components of the MM are plotted in Fig. \ref{TbMnO3}d, which are proportional to the electromagnon contribution to the $\alpha$ and $\alpha'$ tensors.   This shows unambiguously that a partitioning of the electromagnetic response into only the $3 \times 3$ diagonal blocks of the $6 \times 6$ Berreman matrix is incomplete.   One must include the off-diagonal $\alpha$ blocks as well.   Such measurements of the complete electromagnetic response are only possible within a generalized ellipsometry or Mueller matrix approach.

\subsection{Time-domain THz ellipsometry}

An extremely powerful method to characterize materials in the THz range is with THz time-domain spectroscopy (THz-TDS) \cite{Grischkowsky_Oct1990}.  However, as discussed above, since this technique is normally used in transmission mode, this technique falls short in characterization of many metals, thick or highly doped semiconductors, coatings on thick substrates, substances in aqueous solution, and any otherwise opaque compound.  Moreover, an outstanding technical problem with THz-TDS continues to be the determination of absolute spectral values when performing measurements in a reflection geometry  \cite{Nashima_Dec2001}-\cite{Pashkin_2003}. Combining ellipsometry with THz-TDS leads to a new technique called terahertz time-domain spectroscopic ellipsometry (THz-TDSE) that in principle can overcome the shortcomings of conventional THz-TDS. In THz-TDSE, a (sub)picosecond pulse with known polarization state is used as a probe to illuminate the sample, and then the modified polarization state by the sample is detected upon reflection or transmission. Unlike conventional optical ellipsometry, the reflected(transmitted) signal is detected coherently in the time-domain which allows one to obtain both amplitude and phase of the light in the two orthogonal directions. By transforming the time-domain data into the frequency domain through Fourier analysis, it is possible to extract ellipsometric parameter spectra similar to the standard optical spectroscopic ellipsometry.  However, it should be noted that because the instrumentation, signal analysis and calibration methods in THz-TDSE differ from those in the standard optical ellipsometry, all need to be revised.  For instance, in conventional ellipsometry the analyzer polarizer must typically be swept over 360$^\circ$.   However because phase is determined directly in THz-TDS, in THz-TDSE it is only necessary to measure two orthogonal linear polarizations to completely characterize the polarization state.    This is easily done in THz-TDSE with linear polarizers, albeit with possibly significant errors due to polarizer leakage.   We will see below that calibration of such errors is particularly essential in this technique to get accurate data.

\begin{figure}
\centering
  \includegraphics[width=6 cm]{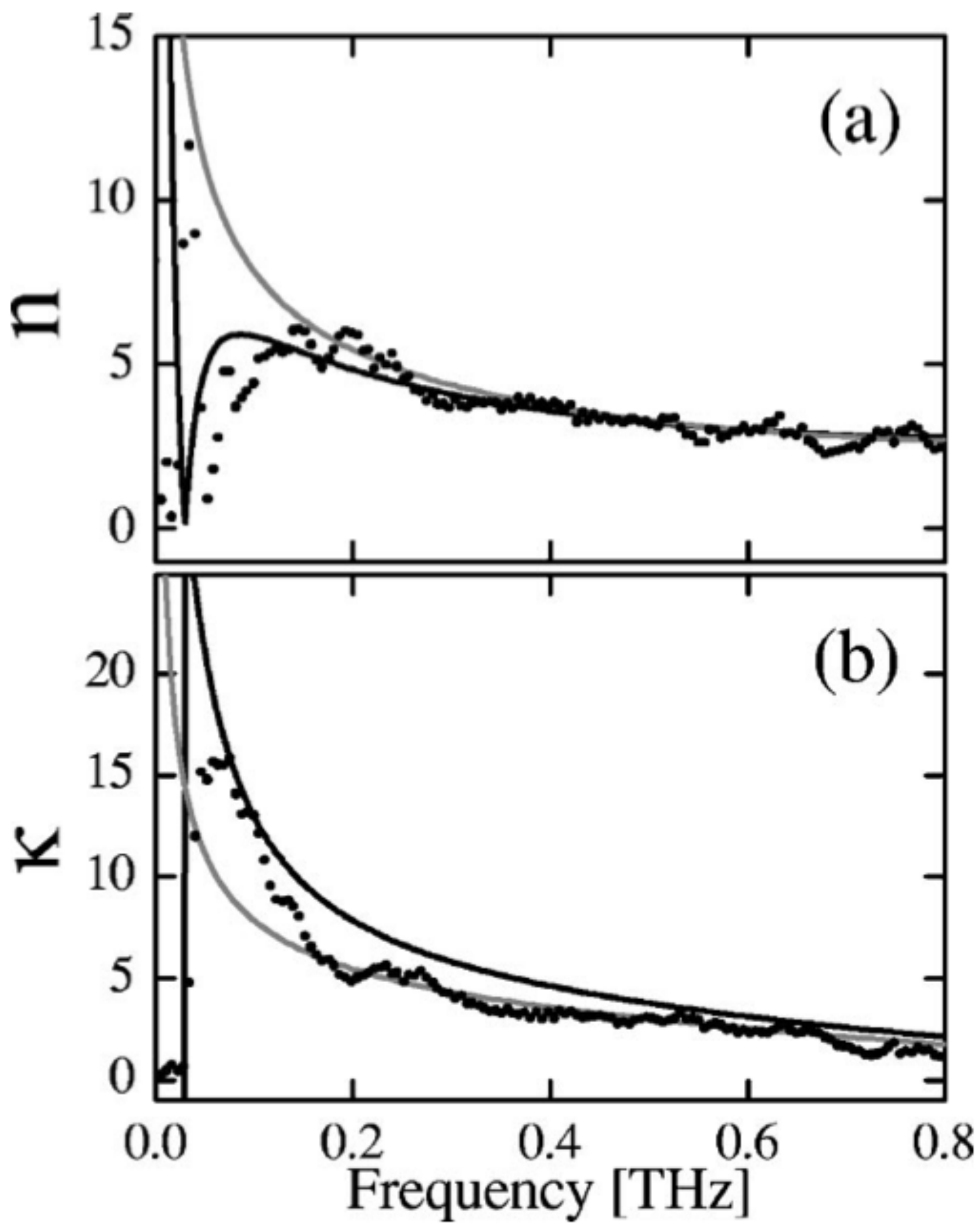}
\caption{Frequency dependence of the real (a) and the imaginary (b) parts of the complex refractive index of a P-doped Si with an error of $\delta \theta = 0.7^\circ$ (solid line) in the rotation angle of the polarizer. For comparison, values without the error of the rotation angle (gray line) and the experimental values (dots) are also shown.  From Ref. \cite{Nagashima_2001}.}
\label{Nagashima}       
\end{figure}

Nagashima and Hangyo made the first attempt at THz-TDSE using a conventional TDS system modified to work at 45$^\circ$ glancing incidence \cite{Nagashima_2001}.   They deduced the complex index of refraction of a P-doped Si wafer from the measurements of the wave forms of reflected s- and p-polarized THz pulses without reference measurement.  The obtained refractive indices above 0.2 THz shows reasonable agreement with that predicted by the Drude model.   Supporting what is known from conventional ellipsometry, they found that even slight misalignments of the rotation angle of the polarizer and analyzer results in a large error of the complex refractive index.  As shown in Fig. \ref{Nagashima}, they estimated that they had an approximately $0.7^\circ$ misaligment of the analyzer polarizer.   Although this introduces only small errors at high frequency, it gives an almost 100\% scale error on the low frequency side.   As will be shown in Sec. \ref{VASE}, these kinds of errors can be compensated through an error correction routine.

Although they referred to it as THz mangeto-optical Kerr effect (MOKE) spectroscopy Shimano \textit{et al.} performed what was essentially time-domain THz ellipsometry where they measured the Kerr rotation and ellipticity of THz light reflected at 45$^\circ$ in the region from 0.5 to 2.5 THz using a THz polarization-sensitive technique \cite{Shimano02a}.  At a static magnetic field of 0.48 T, a large Kerr rotation over 10$^\circ$ originating from magneto-optical resonance is observed in an n-type undoped InAs wafer at room temperature. Later experiments provided an even more complete picture \cite{Ino04a}.

\begin{figure}
\centering
  \includegraphics[width=8 cm]{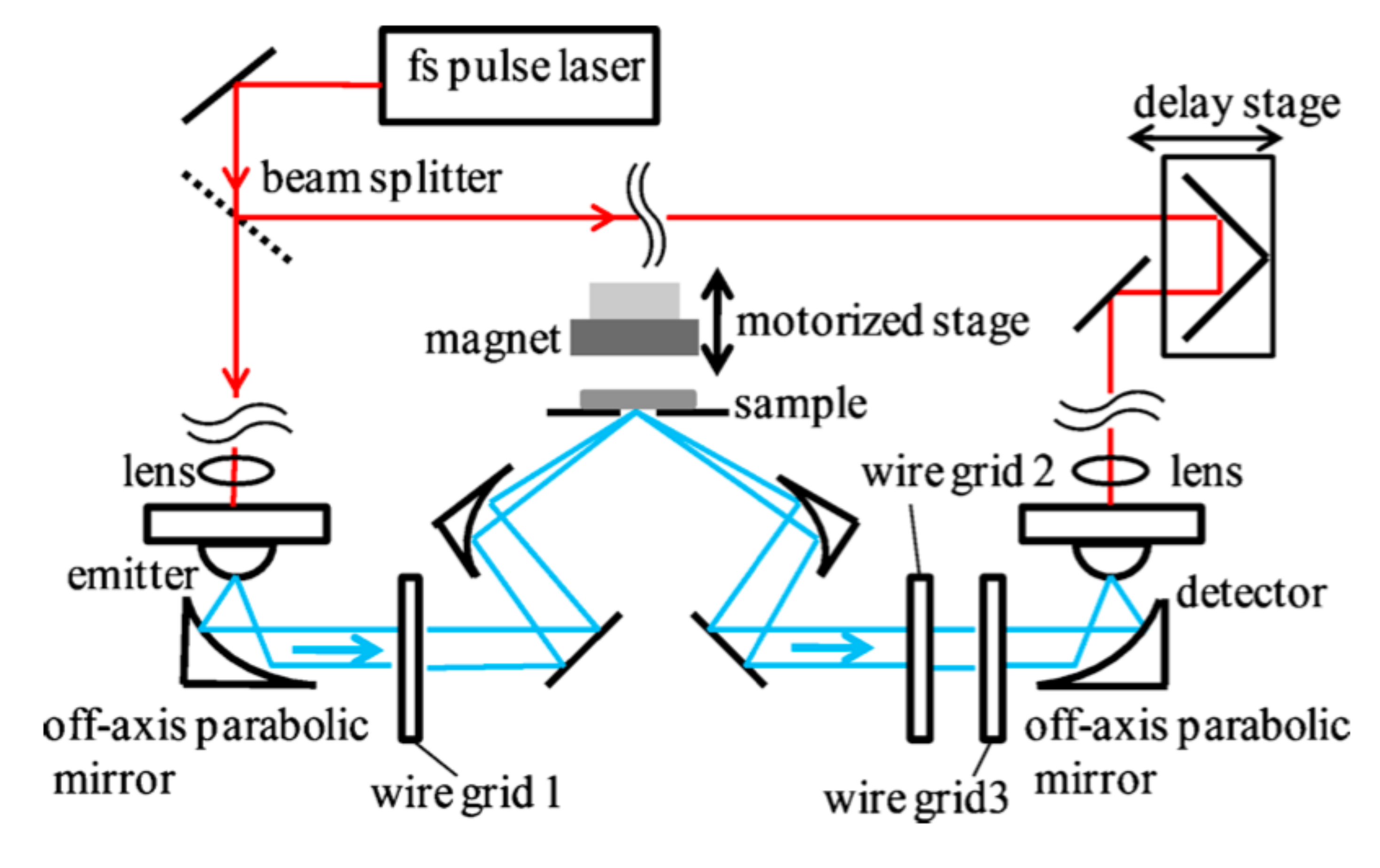}
\caption{Experimental setup for the THz-TDSE of Yatsugi \textit{et al.}.   From Ref. \cite{Yatsugi09a}. }
\label{matsumoto}       
\end{figure}

The Osaka group has built a general use ellipsometer based on THz-TDS capable of a variable angle of incidence (Fig.  \ref{matsumoto}).   THz pulses are focused onto the sample by an off-axis parabolic mirror with an effective focal length of 152 mm. The beam spot size at the focal point is $\sim$ 5 mm in diameter.  In order to adjust the incident angle, the arms are rotated around the center of the sample surface, while the flat mirrors are rotated and translated such that the THz beams are directed onto the off-axis parabolic mirrors.  The time-domain pulses of the $p-$ and $s-$polarized light are measured by changing the angular position of a wire-grid polarizer (B) by 90$^\circ$.  A magnetic field of up to 0.46 T can be applied by a NdFeB permanent magnet placed behind the sample.  Calibration was performed by adjusting the angular positions of the polarizers using an Au mirror as a standard references that has ellipsometric parameters of tan $\Psi \sim 1$ and $\Delta \sim 0$ in the THz frequency region at typical angles of incidence.  The origin of the angular position of the polarizer (B) is adjusted so that the detected $p-$ and $s-$polarized electric fields are strictly equal, although according to our experience such equalization may not be easy to achieve in practice that emphasizes the need for a more sophisticated calibration scheme.

\begin{figure}
\centering
  \includegraphics[width=6 cm]{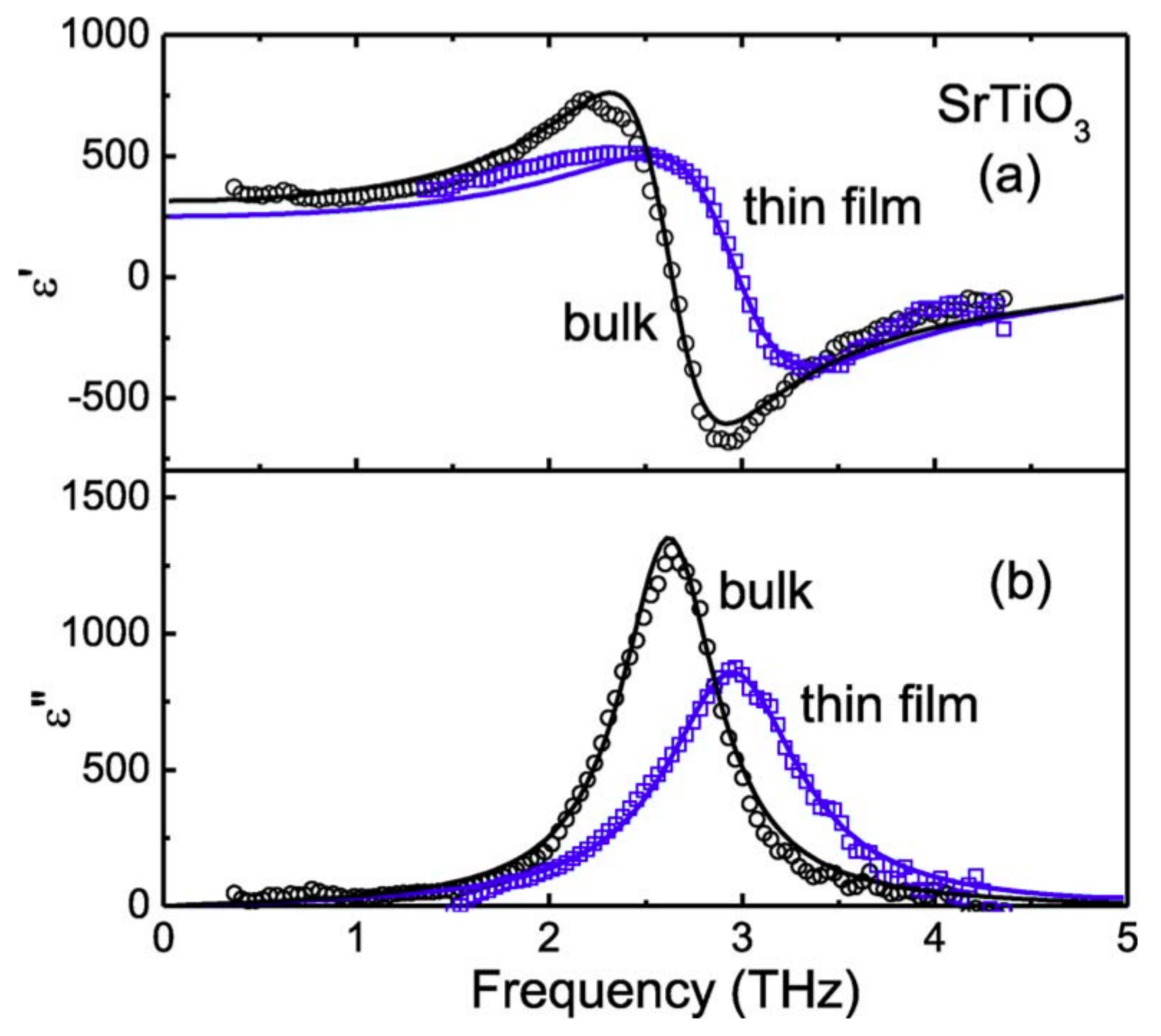}
\caption{(a) Real and (b) imaginary parts of the dielectric constants for thin film of SrTiO$_3$ measured by THz-TDSE as compared to standard bulk STO. The solid curves show calculation results for a generalized harmonic oscillator model. From Ref. \cite{Matsumoto11a} }
\label{STO}       
\end{figure}

\begin{figure}
\centering
  \includegraphics[width=6 cm]{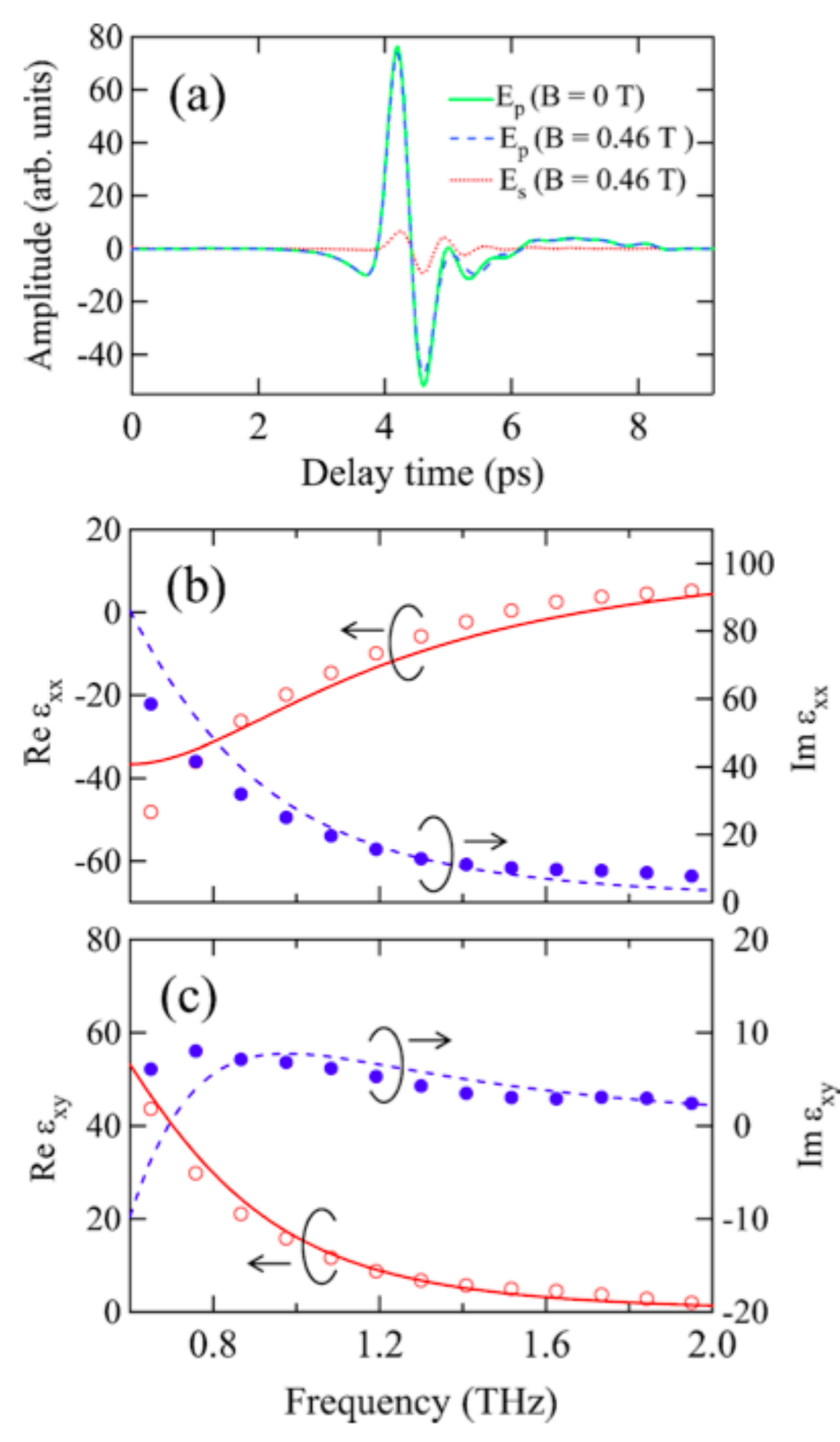}
\caption{Waveforms of THz pulses reflected from the surface of a n-type InAs wafer.  (a) Dashed and solid curves show waveforms of the $p-$polarized waves with and without magnetic field, respectively.  The dotted curve represents the waveform of the $s-$polarized pulses with B=0.46 T.  (b) Diagonal and (c) off-diagonal components of complex dielectric tensor. Open and solid circles indicate the real and imaginary parts of the components, respectively. Solid and dashed curves show Drude model predictions.  From Ref. \cite{Yatsugi09a}. }
\label{Yatsugi09a}       
\end{figure}

This group has demonstrated the potential of THz-TDSE by measuring the complex optical constants of a Si wafer with low resistivity, the soft-mode dispersion of SrTiO$_3$ bulk single crystals, the dielectric constants of doped GaAs thin films, and the Drude parameters of a n-type InAs wafer \cite{Nagashima_2001,Matsumoto_2009,Matsumoto11a,Yatsugi09a}.  Shown in Fig. \ref{STO} is the real and imaginary parts of the dielectric constant SrTiO$_3$ films  obtained at an incident angle of 79.5$^\circ$.   A large prominent soft-mode transverse optical phonon is found at approximately 3 THz.   As can be seen from Fig. \ref{STO} this phonon shows a slight hardening with respect to the bulk case.   A magnetic field allowed complex diagonal and off-diagonal dielectric constants and the complete Drude parameters of band mass, carrier density and scattering rate to be measured in n-type InAs as shown in Fig. \ref{Yatsugi09a}.

\section{Variable Angle THz Time-Domain Spectroscopic Ellipsometry (THz-TDSE)}
\label{VASE}

Until recently, none of the proposed THz-TDSE experimental setups could provide an easy way of changing the angle of incidence without tedious work of optical/terahertz beam realignments. In a recent work \cite{Neshat_Dec2012}, the present authors have developed a setup for THz-TDSE in which the incidence angle in the reflection mode can change very easily in the range of $15^\circ-85^\circ$ with no need for any realignment. The same setup can be transformed into transmission mode with the same ease.

\begin{figure}[htbp]
\centering
\subfigure[]{
   \includegraphics[width=0.75\textwidth] {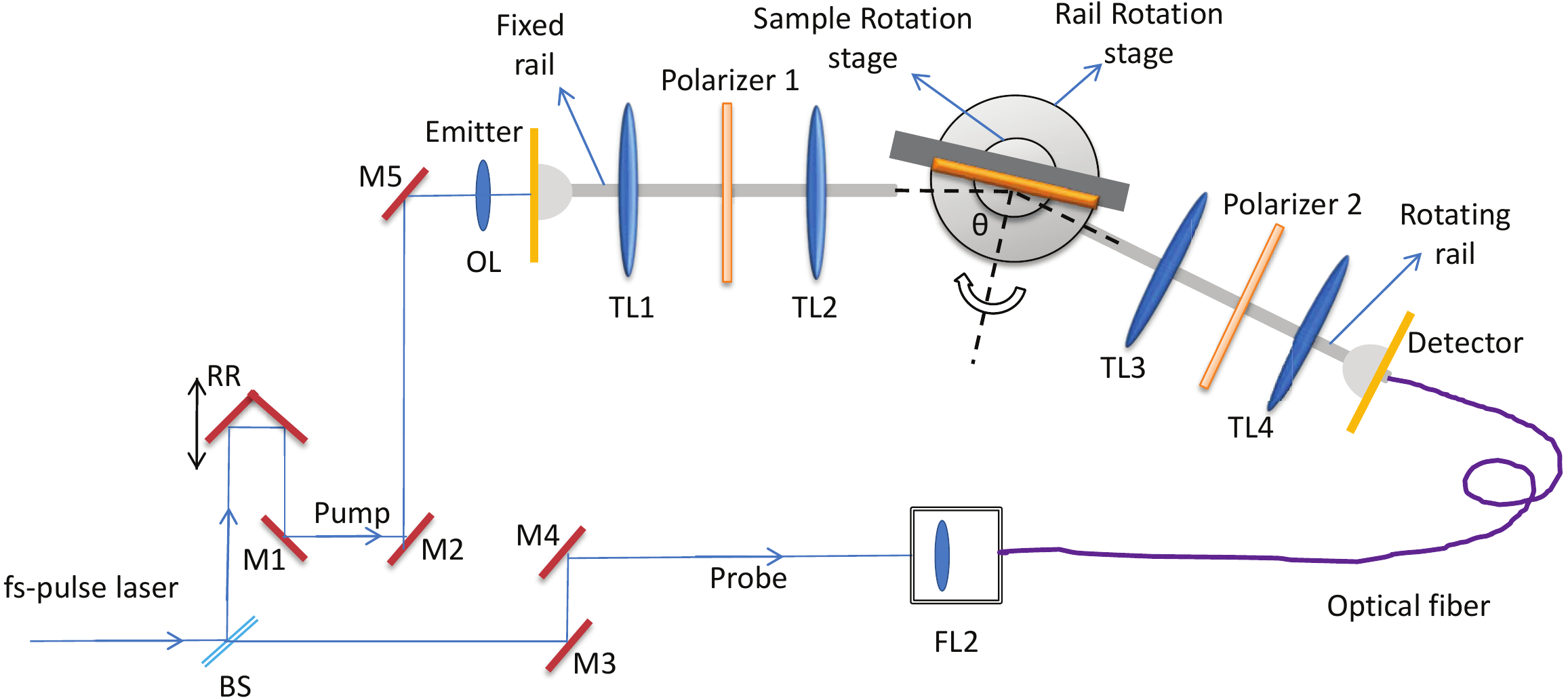}
}\\
\subfigure[]{
   \includegraphics[width=0.25\textwidth] {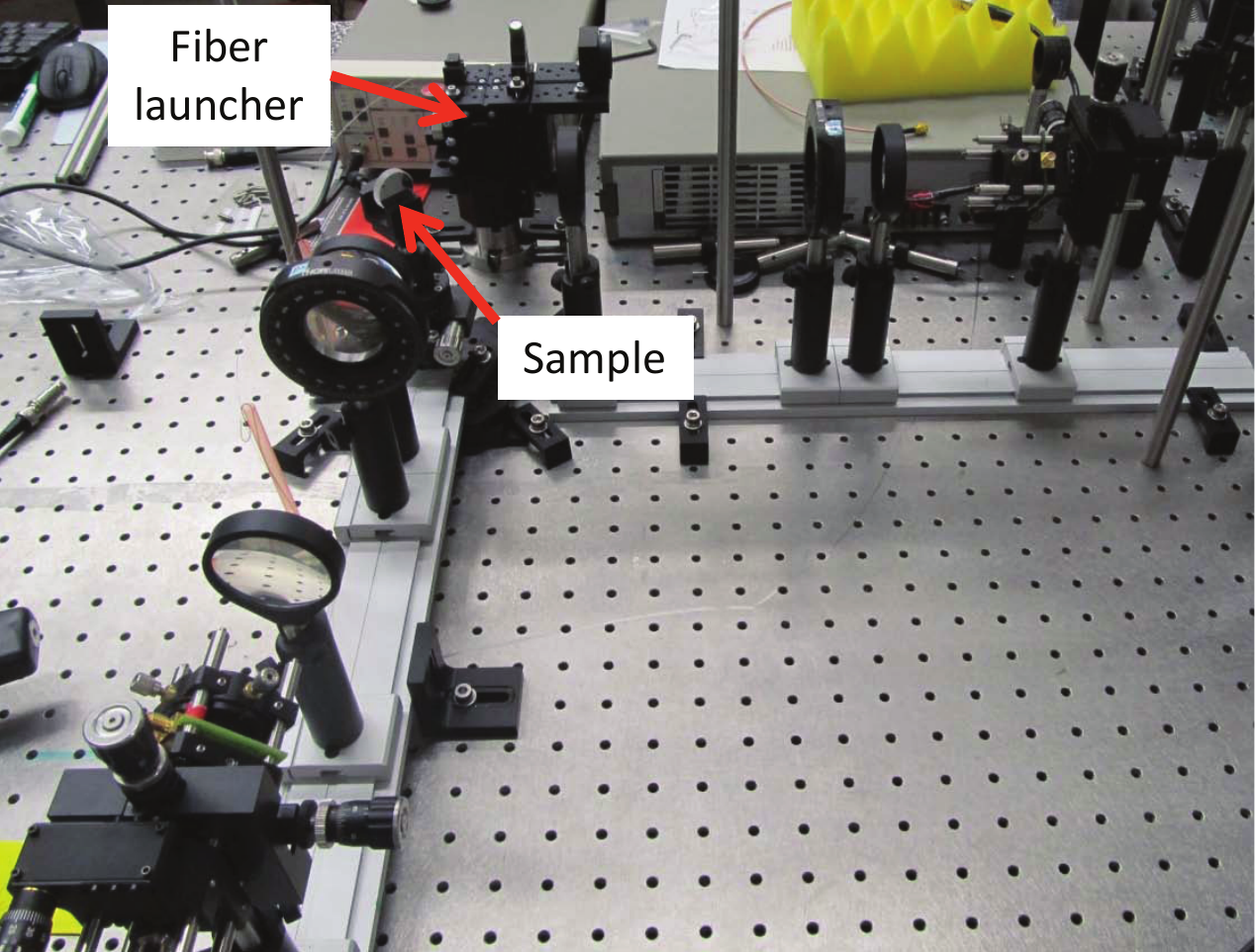}
}
\subfigure[]{
   \includegraphics[width=0.5\textwidth] {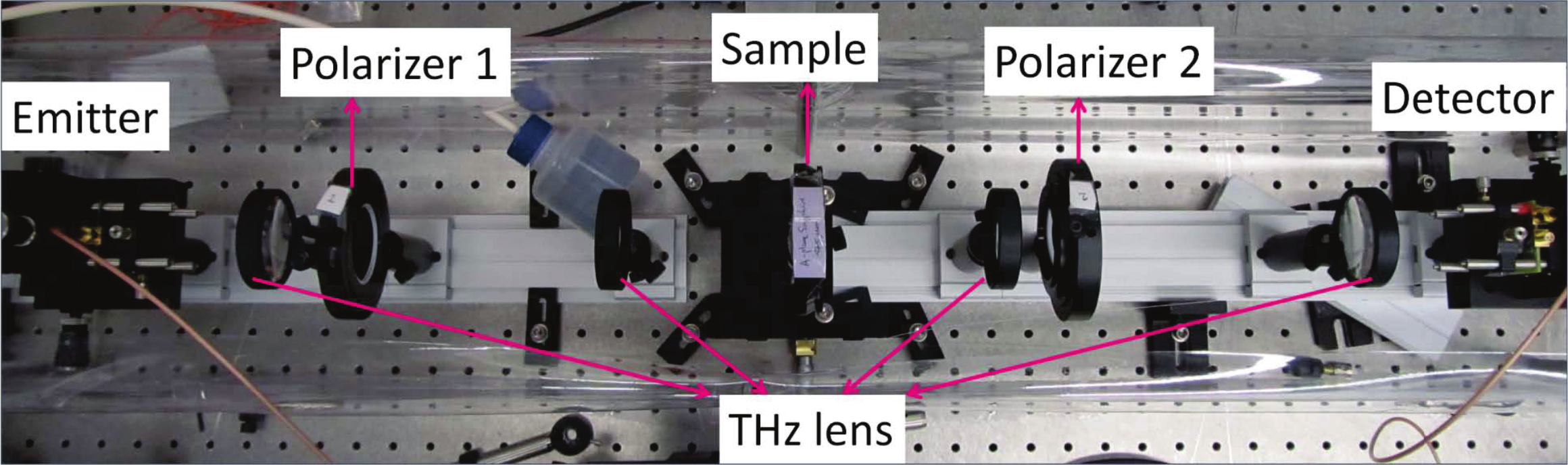}
}
\caption{(a) Schematic of the THz-TDSE setup with variable incidence angle in reflection mode that is also configurable in transmission mode (M, mirror; RR, retro-reflector; BS, optical beam splitter; OL, optical lens; TL, terahertz lens), (b) Lab setup configured in reflection mode at $45^\circ$ incidence angle and (c) transmission mode. From Ref. \cite{Neshat_Dec2012}}\label{Setup}
\end{figure}

Figure \ref{Setup} illustrates our terahertz time-domain spectroscopic ellipsometry setup. In this system, the terahertz components are arranged on two straight arms using optical rails; one arm with THz emitter is fixed on the optical table, whereas the other arm with THz detector can rotate around a center point where the sample is placed. Laser light is coupled into the moving detector using a fiber optic cable.  The sample sits on a second rotation stage. Such a configuration provides variable incidence angle \mbox{($15^{\circ}<\theta<85^\circ$)} in reflection mode, and can be easily configurable in transmission mode by aligning the arms along a straight line (Fig. \ref{Setup}(c)).

The setup uses an $8f$ confocal geometry with terahertz lenses made of \mbox{poly-4-methyl-pentacene-1} (TPX), which is a terahertz- and optically transparent material. Terahertz lenses have 50.8 mm clear aperture diameter and 100 mm focal length, and are less prone to misalignments and polarization distortion as compared to off-axis parabolic mirrors. The terahertz beam profile can be approximated as Gaussian. The sample is placed at the focal point of the terahertz lenses where an effective flat phase front exists at the beam waist.  As discussed above, the strong focusing and long wavelengths minimize the effects of a spread of incidence angles because an effective flat phase front exists near the sample position.  This is the opposite approach than has been used in previous infrared ellipsometers that were designed to work at high $f$-number \cite{Kircher97a}. Two identical photoconductive dipole antennas with collimating substrate lens are used as THz emitter and detector. Two rotatable polarizers are placed in the collimated beams immediately before the detector, and after the emitter as shown in Fig. \ref{Setup}. Polarizers were wire grid with wire diameter and spacing of $10~\mu \textrm{m}$ and $25~\mu \textrm{m}$, respectively, and field extinction ratio of $\sim 40:1$ at 1 THz.

The laser source is an 800 nm Ti:Sapphire femtosecond laser with pulse duration of \mbox{$<20$ fs} and 85 MHz repetition rate, which is divided into pump and probe beams. The pump beam is guided and focused onto the gap of the emitter photoconductive antenna through free-space optics in the conventional fashion. However, the probe beam is guided toward the detector antenna on the rotating arm by a 1 m long hollow core photonic band-gap fiber (HC-PBGF). Using a fiber-coupled THz detector facilitates the rotation of the detector arm with no need for realignment of the optics for each incidence angle. The advantage of HC-PBGF is that it  exhibits extremely low nonlinearity, high breakdown threshold, zero dispersion at the design wavelength, and negligible interface reflection \cite{Hensley_2008}.  The near zero dispersion of such fibers eliminates the need for optical pulse pre-chirping, which makes the optical setup and the alignments considerably simpler than using conventional fibers where pre-chirping is a necessity.

\begin{figure}[htbp]
\centering
\subfigure[]{
   \includegraphics[width=0.375\textwidth] {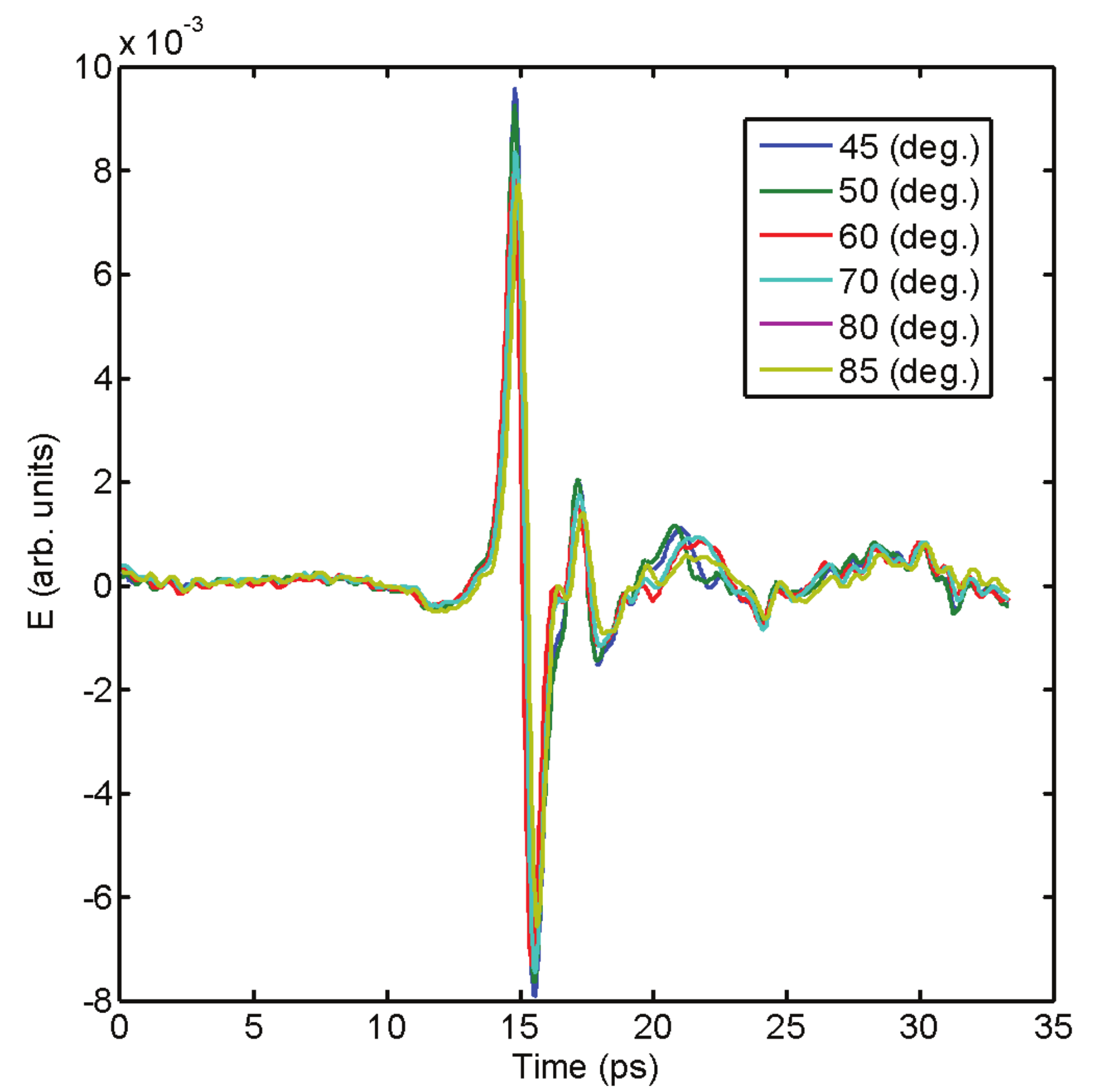}
}
\subfigure[]{
 \includegraphics[width=0.375\textwidth] {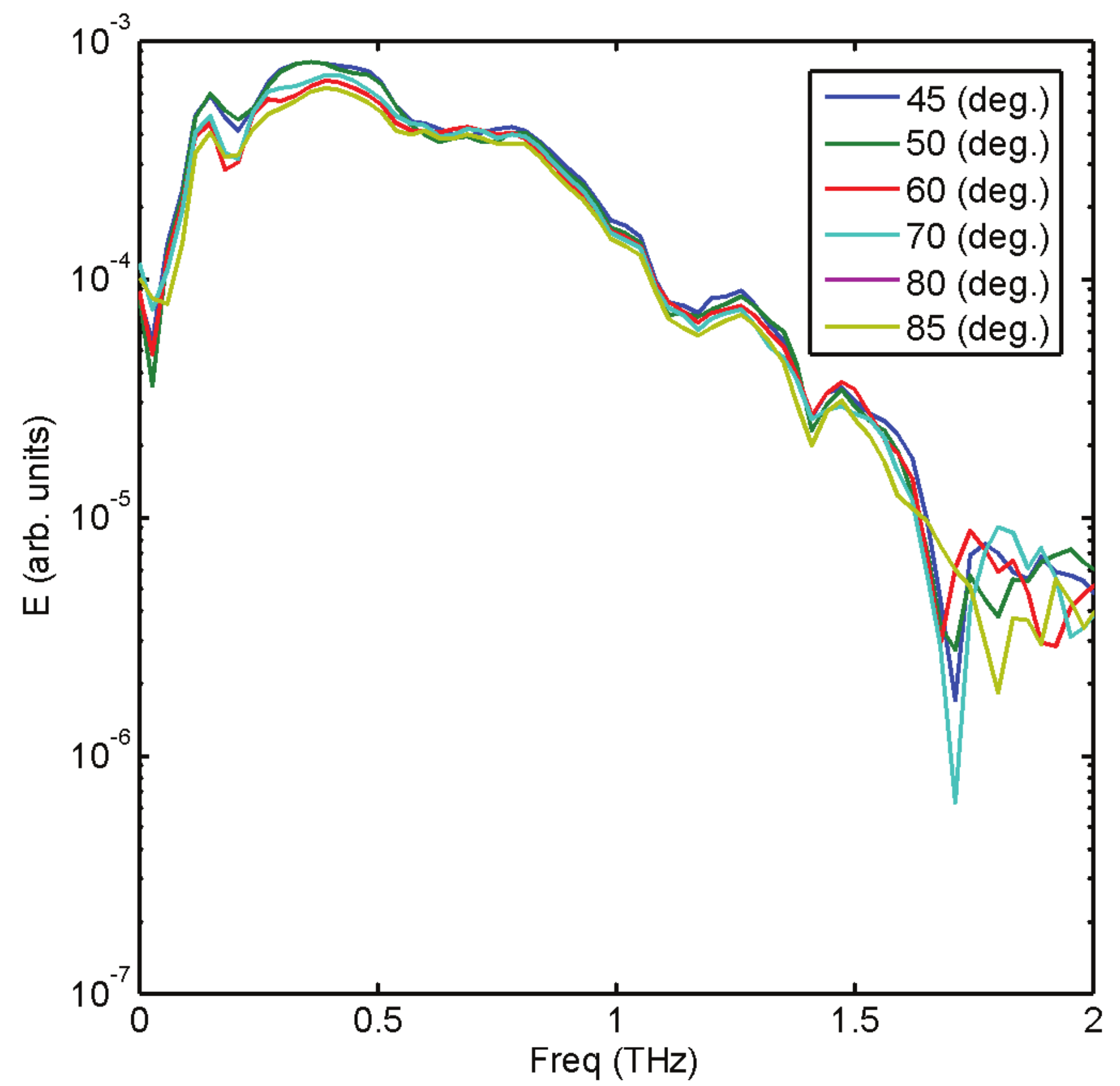}
}
\caption{(a) Time-domain THz pulse and (b) its corresponding spectrum in reflection mode for various incidence angles. A silver mirror was used as the reflector. From Ref. \cite{Neshat_Dec2012}}\label{PulsReflection}
\end{figure}

The rest of the setup is conventional for THz-TDS.  The temporal terahertz pulse is acquired by a lock-in amplifier during a time window within which the time delay between terahertz pulse and the sampling probe laser is swept by continuous movement of the retro-reflector. The pump beam is mechanically chopped, and the lock-in amplifier is synchronized to the chopping frequency. The acquired temporal signal is then taken into the frequency domain through a Fourier transform. Figure \ref{PulsReflection} shows the THz pulse and its corresponding spectrum in reflection using fiber coupling for various incidence angles without any optics realignment after changing the incidence angle. One can see that in this setup, any changes in the overall intensity at different angles are very small. In the reflection measurements a silver mirror was used as a reflector.  In order to avoid water vapor absorption, the space with terahertz wave propagation was enclosed and purged with dry air during the measurements.  The polarimetry is accomplished by alternately measuring with polarizer 2 in the $0^\circ$ and $90^\circ$ positions with polarizer 1 in the $45^\circ$ position.

Until recently, little attention had been paid to the compensation through calibration schemes of non-idealities in optical components and their alignment in THz-TDSE.  Such compensation and calibration is essential in conventional optical range ellipsometry for accurate measurements, and as shown in \cite{Neshat_Dec2012} is no-less essential in THz-TDSE. In \cite{Neshat_Dec2012}, the authors proposed a calibration scheme that can compensate for the non-ideality of the polarization response of the THz photoconductive antenna detector as well as that of wire grid polarizers used in the setup. In this calibration scheme, the ellipsometric parameters are obtained through a regression algorithm that has been adapted from the conventional regression calibration method developed for rotating element optical ellipsometers \cite{Johs_1993}. As a demonstration, the characterization for a high resistivity silicon substrate and a highly P-doped Si substrate along with the capacity to measure a few micron thick grown thermal oxide on top of Si have been presented. 

\begin{figure}[htbp]
\centering
\includegraphics[width=5.5cm] {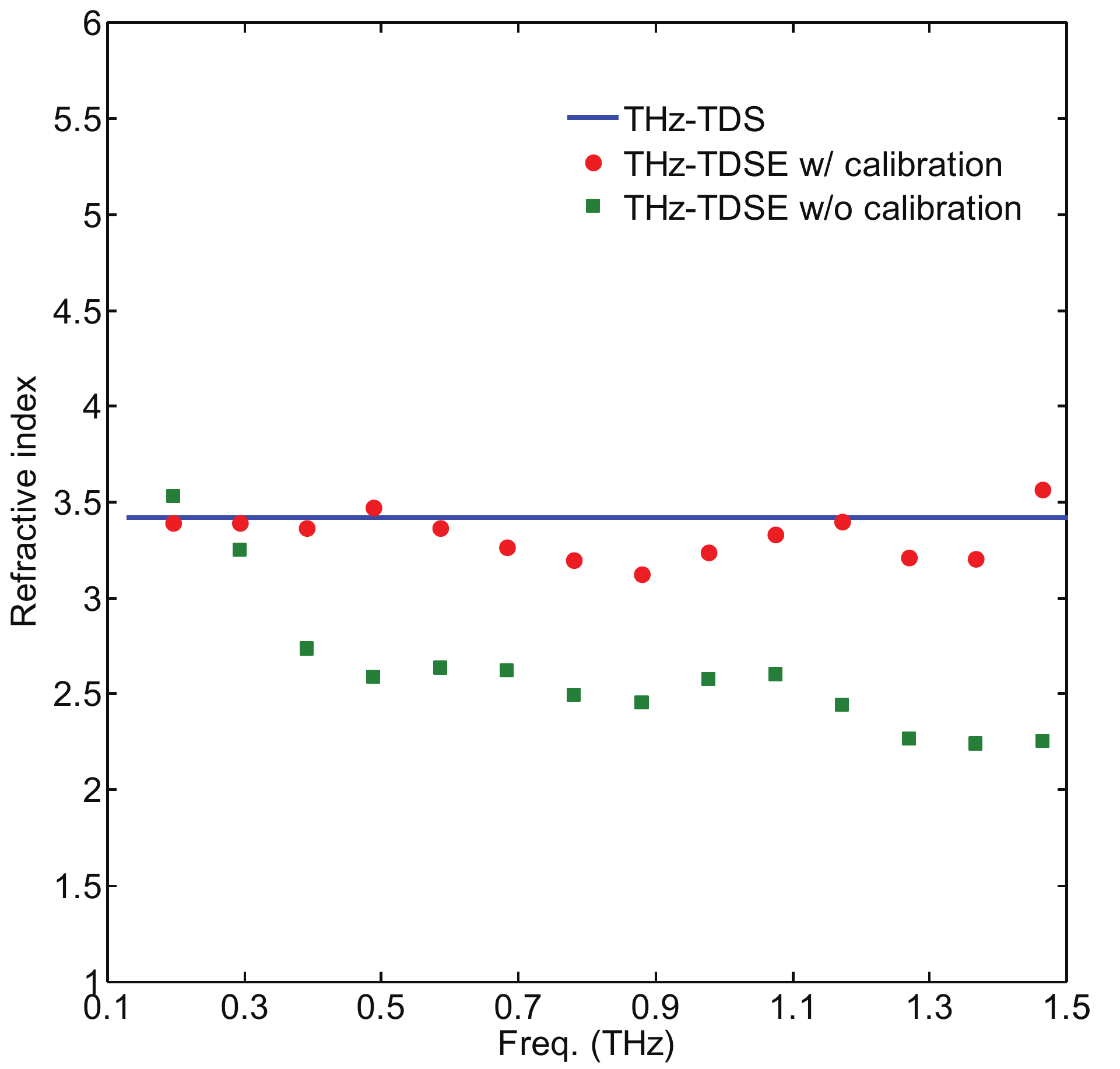}
\caption{Extracted refractive index of the high resistivity silicon substrate from the uncalibrated (square) and calibrated (circle) ellipsometric parameters. Solid line shows the data from the conventional transmission THz-TDS reported in \cite{Grischkowsky_Oct1990}. From Ref. \cite{Neshat_Dec2012}}\label{RI_SiHR}
\end{figure}

Figure \ref{RI_SiHR} compares the extracted refractive index of a high resistivity silicon substrate from the uncalibrated and calibrated ellipsometric parameters with that from the conventional transmission THz-TDS reported in \cite{Grischkowsky_Oct1990}. It is evident that the calibration has substantially improved the measurement accuracy over the uncalibrated data.   The ellipsometric measurements do not have as much precision as the transmission experiments. But this illustrates the important point that the power of ellipsometric measurements rests not on measuring dielectric properties of samples that could be measured in transmission, but in measuring samples that are otherwise opaque.

\begin{figure}[htbp]
\centering
   \includegraphics[width=5.5cm] {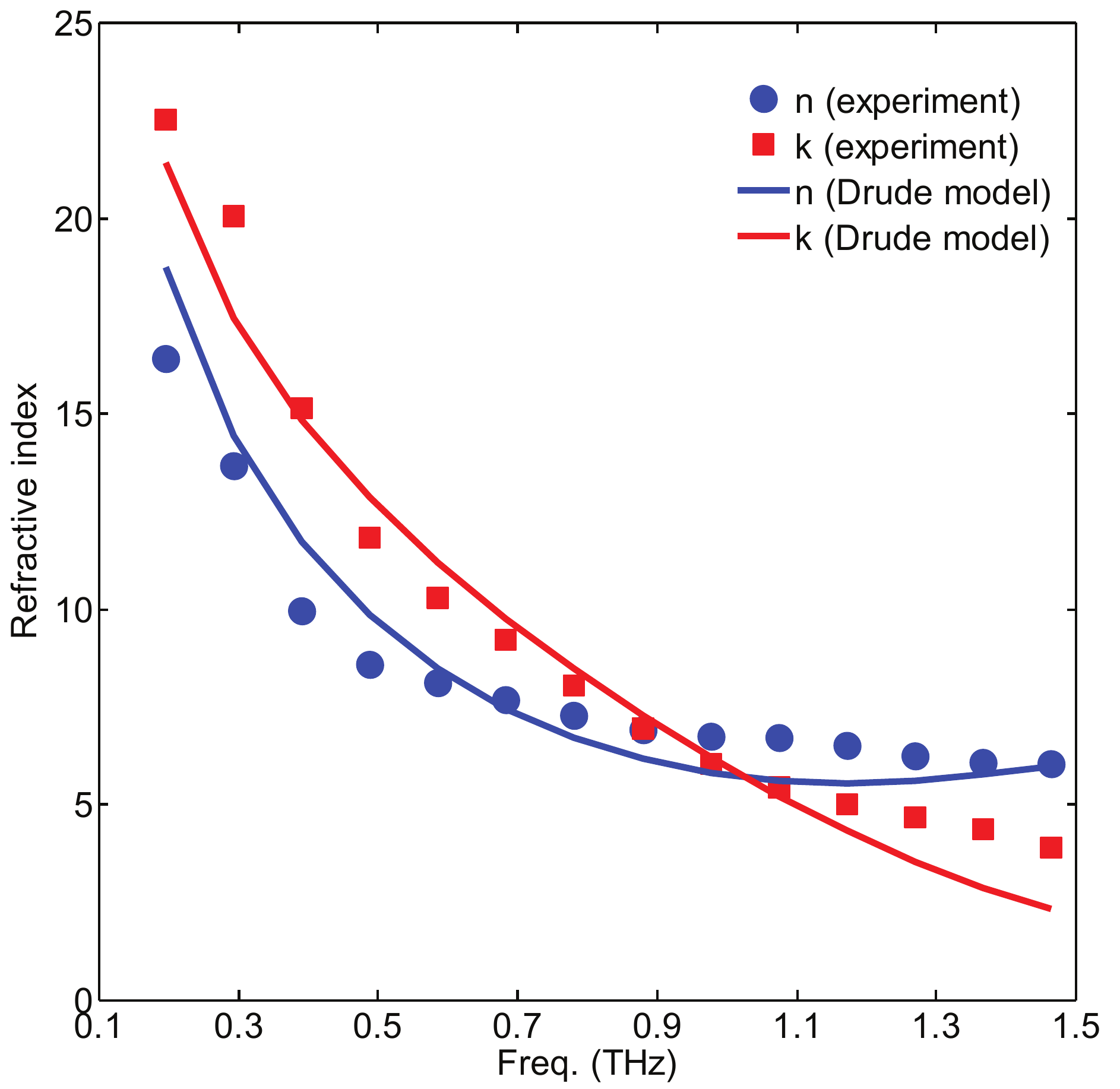}
   \includegraphics[width=5.7cm] {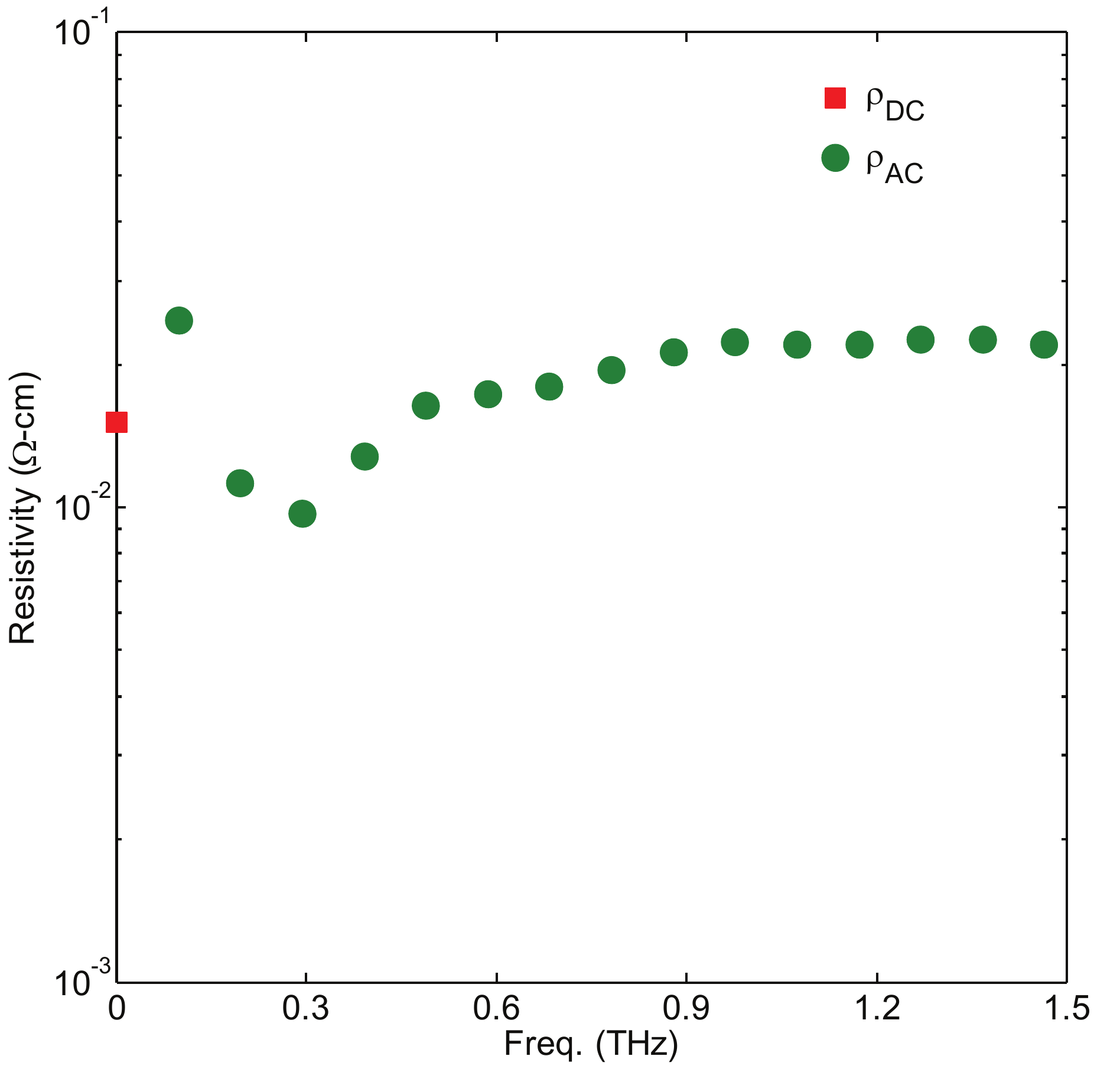}
\caption{(a) Extracted complex refractive index of the highly doped silicon substrate from ellipsometric parameters. Solid lines show the Drude model fit to the measured data.  (b) Extracted resistivity of the highly doped silicon substrate from ellipsometry measurement. Square mark presents its DC resistivity measured via a non-contact eddy-current resistivity gauge (COTS ADE 6035). From Ref. \cite{Neshat_Dec2012}}\label{Si0151}
\end{figure}

Figure \ref{Si0151}(a) shows the extracted complex refractive index of the highly doped silicon substrate from ellipsometric parameters along with a Drude model fit to the measured data. For the substrate resistivity and scattering time we obtain $\rho=0.011~\Omega$-cm and $\tau=190$ fs from the fitting and by assuming an effective electron mass of 0.26m$_0$ \cite{Martin_1990}. These values are reasonable when compared with those previously reported in \cite{Hofmann10a} for a highly doped Si substrate. Figure \ref{Si0151}(b) shows the AC resistivity of the highly doped silicon substrate extracted from the ellipsometry measurement. As shown, the measured AC resistivity is consistent with the DC resistivity measured via a non-contact eddy-current resistivity gauge (COTS ADE 6035). However, there is a dip in the resistivity curve around 0.3 THz. The exact reason of the dip is not clear to us at present, but we realized that polishing the substrate surface has considerable effect on its depth. Therefore, some surface effect may cause such feature in the resistivity curve.

\begin{figure}[htbp]
\centering
\subfigure[]{
 \includegraphics[width=0.375\textwidth] {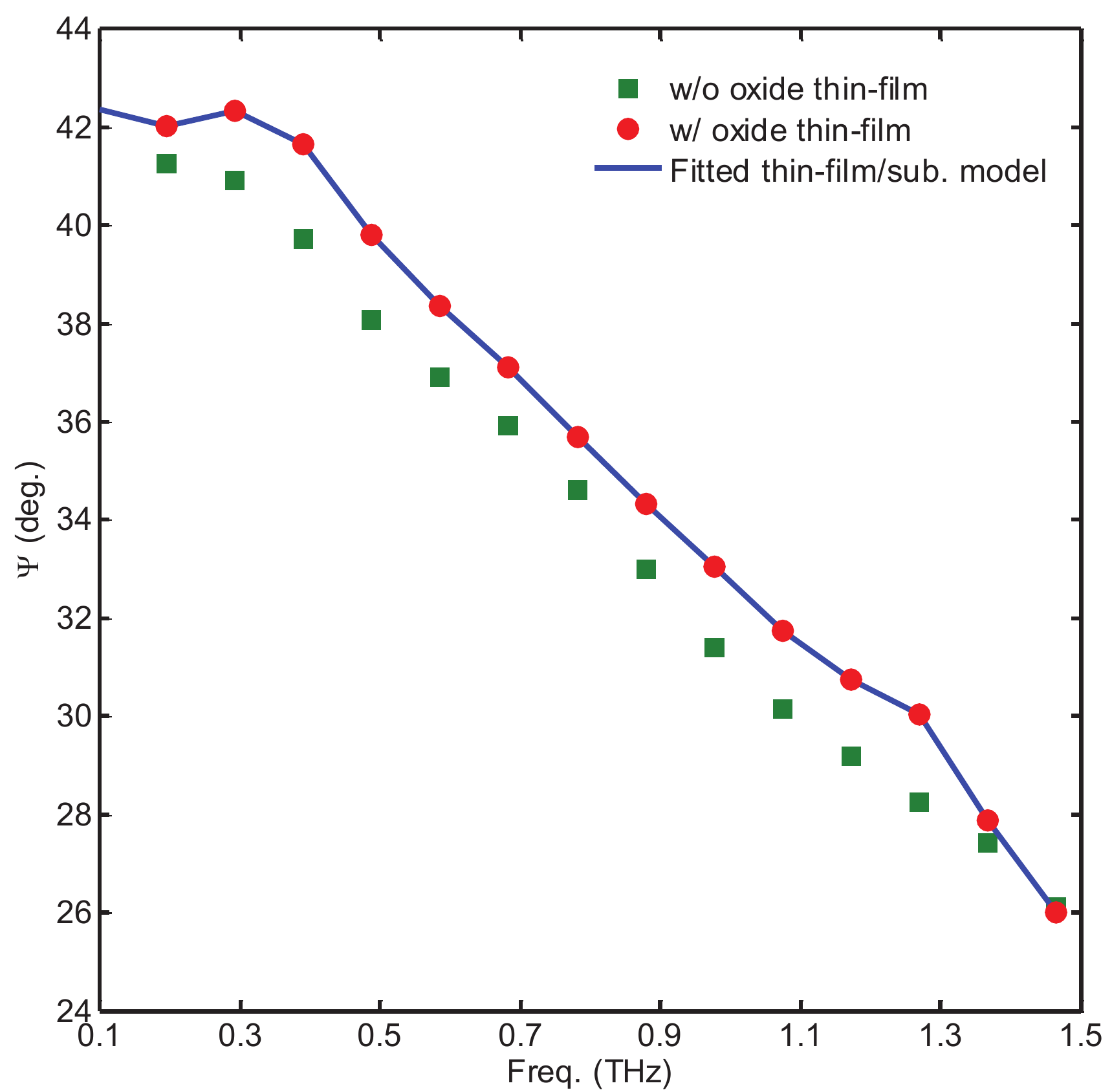}
}
\subfigure[]{
  \includegraphics[width=0.375\textwidth] {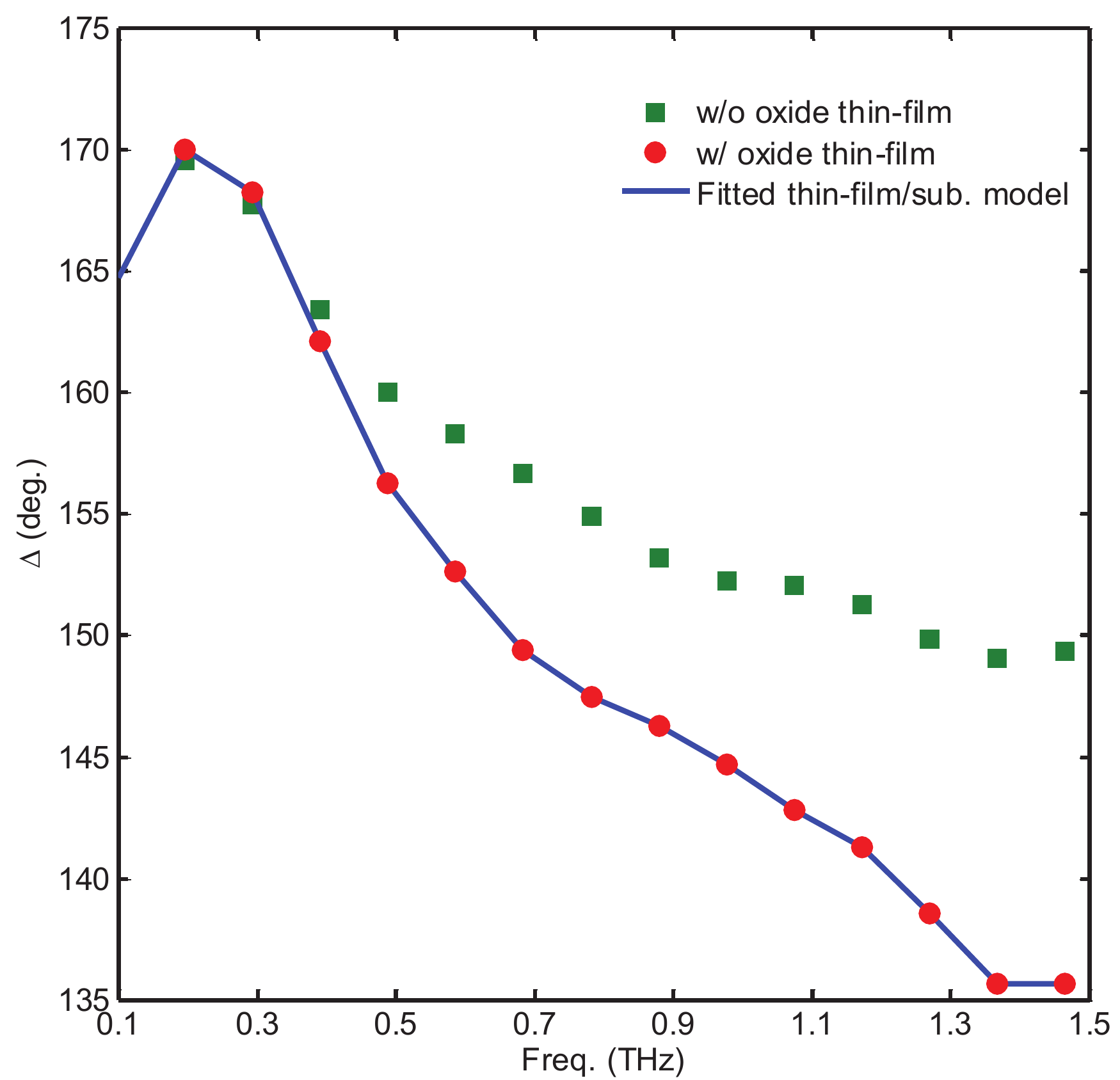}
}
\caption{Comparison between measured ellipsometric parameters (a) $\Psi$ and (b) $\Delta$ for the highly doped Si substrate w/ and w/o oxide thin-film. Solid line shows fitted thin-film/substrate model. The incidence angle was $73^\circ$. From Ref. \cite{Neshat_Dec2012}}\label{oxide}
\end{figure}

In order to further verify the overall accuracy and utility of our THz-TDSE measurement method, we grew thermally a thin layer of oxide on top of the highly doped Si substrate. The thickness of the oxide thin-film was measured independently as 1.95 $\mu$m using optical techniques. Figure \ref{oxide} compares the measured ellipsometric parameters of the highly doped Si substrate before and after oxidization. The thickness of the oxide thin-film was extracted by fitting a \mbox{thin-film/substrate} model to the measured ellipsometric parameters. We used the refractive index taken from Fig. \ref{Si0151}(a) for the substrate, and considered the refractive index of the thin-film as well as its thickness as free parameters in the fitting process. The resultant thickness of the oxide thin-film was obtained as 1.9 $\mu$m that is in excellent agreement with that obtained from the independent optical method. The fact that we can measure dielectric thicknesses that are only 0.3\% of the free space wavelength of 500 GHz radiation further shows the high accuracy of the technique and methodology.   In ongoing work we have demonstrated that one can use the THz-TDSE technique to actually measure the thickness of dielectric layers buried under optically opaque metallic conducting layers \cite{Neshat13a}.  This may be useful in semiconductor applications where one needs characterize a layered structure in electro-optical devices.   This instrument is now also being used in a number of projects to characterize materials with interesting ``correlated" electronic properties such as the temperature dependent metal-insulator transition in V$_2$O$_3$ \cite{Wu13a}.

\section{Outlook}
\label{sec:4}

THz ellipsometry has seen dramatic advances in recent years.   The technique has gone from simple ``proof-of-principle" style demonstrations to its use as a real tool for the characterization of exotic electromagnetic effects in solids.  Looking forward one may expect continued technological advances and the continued application of THz ellipsometry to many more interesting materials systems.   It will be interesting to couple the methodologies developed for THz ellipsometry to recent developments for high precision THz polarimetry \cite{Aguilar,Castro05a,Morris12a,George12a,Aschaffenburg12a}.   Some of these methods can achieve reliable measurements of the polarization state of THz light with precision as high as 0.02$0^\circ$.  New calibration methods will have to be developed, but these high accuracies will presumably lead to dramatic advances as well.  New analysis techniques and configurations will be developed as well as the extension of existing methods to extreme conditions (high magnetic field and low temperatures) to explore interesting states of matter in a fashion simply not possible previously.

\begin{acknowledgements}

This work was made possible by support from the Gordon and Betty Moore Foundation through Grant GBMF2628 to NPA and DARPA YFA N66001-10-1-4017.

\end{acknowledgements}

\bibliographystyle{spphys}       

\end{document}